%% file: JonesSilicates2012_final_ms.tex
\documentclass[useAMS, usenatbib]{mn2e}

\usepackage{deluxetable}
\usepackage{amssymb}
\usepackage{graphicx}
\usepackage{threeparttable}
\usepackage{multirow}
\usepackage{longtable}
\usepackage{subfig}
\usepackage{url}

\newcommand{\mum}{\ifmmode{\rm \mu m}\else{$\mu$m }\fi}

\newcommand{\bra}{\ifmmode{\rm Br \alpha}\else{Br $\alpha$}\fi}
\newcommand{\zx}{\ifmmode{\chi}\else{$\chi$}\fi}
\newcommand{\zpp}{\ifmmode{\pi^1}\else{$\pi^1$}\fi}
\newcommand{\zj}{\ifmmode{\theta}\else{$\theta$}\fi}
\newcommand{\zm}{\ifmmode{\mu}\else{$\mu$}\fi}
\newcommand{\za}{\ifmmode{\alpha}\else{$\alpha$}\fi}

\newcommand{\Msun}{\ensuremath{{\rm M}_{\odot}}}      
\newcommand{\Msuny}{\ensuremath{{\rm M}_{\odot} \, {\rm yr}^{-1}}}    
\newcommand{\chisq}{\ifmmode{\chi^{2} }\else{$\chi^2$}\fi}

\title[Crystalline silicates around O-AGB \& RSG stars] 
{On the metallicity dependence of crystalline silicates in oxygen-rich asymptotic giant branch stars and red supergiants}

\author[O. Jones et al.]   
{O.~C.~Jones,$^1$ \thanks{E-mail:ojones@jb.man.ac.uk}
  F.~Kemper,$^{2}$ 
  B.~A.~Sargent,$^{3,4}$
  I.~McDonald,$^1$
  C.~Gielen,$^{5}$
 \newauthor 
  Paul~M.~Woods,$^{6}$
  G.~C.~Sloan,$^7$
  M.~L.~Boyer,$^{4}$
  A.~A.~Zijlstra,$^1$
  G.~C.~Clayton,$^8$
  \newauthor
  K.~E.~Kraemer,$^{9}$
  S.~Srinivasan$^{10}$
  and P.~M.~E.~Ruffle$^{1}$ \\
$^1$ Jodrell Bank Centre for Astrophysics, Alan Turing Building, Oxford Road, Manchester, M13 9PL, UK.\\
$^{2}$ Academia Sinica Institute of Astronomy and Astrophysics, PO Box 23-141, Taipei 10617, Taiwan.\\
$^{3}$ Center for Imaging Science, Rochester Institute of Technology, 54 Lomb Memorial Drive, Rochester, NY 14623, USA.\\
$^{4}$ Space Telescope Science Institute, 3700 San Martin Drive, Baltimore, MD 21218, USA.\\
$^{5}$ Instituut voor Sterrenkunde, Katholieke Universiteit Leuven, Celestijnenlaan 200D, 3001 Leuven, Belgium. \\ 
$^{6}$ Department of Physics and Astronomy, University College London, Gower Street, London, WC1E 6BT, UK.\\
$^{7}$ Department of Astronomy, Cornell University, Ithaca, NY 14853, USA.\\
$^{8}$ Department of Physics and Astronomy, Louisiana State University, Baton Rouge, LA 70803, USA. \\ 
$^{9}$ Institute for Scientific Research, Boston College, Kenny Cottle L106B, Newton, MA 02459-1161, USA. \\
$^{10}$ UPMC-CNRS UMR7095, Institut d'Astrophysique de Paris, F-75014 Paris, France.}

\begin{document}

\date{Accepted XXXX XXXXXX XX. Received XXXX XXXXXX XX}

\pagerange{\pageref{firstpage}--\pageref{lastpage}} \pubyear{2012}

\maketitle

\label{firstpage}

\begin{abstract}
 We investigate the occurrence of crystalline silicates in oxygen-rich evolved stars across a range of metallicities and mass-loss rates. It has been suggested that the crystalline silicate feature strength increases with increasing mass-loss rate, implying a correlation between lattice structure and wind density. 
To test this, we analyse {\em Spitzer} IRS and {\em Infrared Space Observatory} SWS spectra of 217 oxygen-rich asymptotic giant branch and 98 red supergiants in the Milky Way, the Large and Small Magellanic Clouds and Galactic globular clusters. These encompass a range of spectral morphologies from the spectrally-rich which exhibit a wealth of crystalline and amorphous silicate features to `naked' (dust-free) stars. 
We combine spectroscopic and photometric observations with the {\sc grams} grid of radiative transfer models to derive (dust) mass-loss rates and temperature. We then measure the strength of the crystalline silicate bands at 23, 28 and 33 $\mu$m. We detect crystalline silicates in stars with dust mass-loss rates which span over 3 dex, down to rates of $\sim$10$^{-9}$ M$_\odot$ yr$^{-1}$.
 Detections of crystalline silicates are more prevalent in higher mass-loss rate objects, though the highest mass-loss rate objects do not show the 23-\mum feature, possibly due to the low temperature of the forsterite grains or it may indicate that the 23-\mum band is going into absorption due to high column density. Furthermore, we detect a change in the crystalline silicate mineralogy with metallicity, with enstatite seen increasingly at low metallicity.   
\end{abstract}


\section{Introduction}

Cool evolved stars in the later stages of active nuclear burning expel a large fraction of their mass through a slow, dense stellar wind at rates of $10^{-7}$ to $10^{-4}$ \Msuny \citep{Habingbook}. In low- to intermediate-mass stars (0.8--8~M$_\odot$) this occurs on the asymptotic giant branch (AGB), while higher-mass stars (8--25 \Msun) undergo this intense mass loss as red supergiants (RSG). In AGB stars it is believed that radial pulsations levitate gaseous material above the surface of the star \citep{Habingbook}, resulting in a cooling outflow from which molecular species form and (at temperatures below $\sim$1400 K) dust grains condense \citep{Hoefner1998}. A dust-driven wind develops, propelled by radiation pressure from the stellar photons  \citep{Gehrz1971, Bowen1991, Norris2012}. For RSGs, a slightly different mechanism may drive the dense outflow \citep{JosselinPlez2007}.

Chemistry in AGB atmospheres is affected by mixing of the stellar atmosphere and interior, known as dredge up. This process brings nuclear-fusion products, particularly carbon, to the stellar surface, enhancing their abundances. Chemistry in the cooling wind is dominated by the relative abundances of carbon and oxygen, as these combine at high temperatures to form carbon monoxide. The remaining free carbon or oxygen then determines whether the star produces C-rich dust (such as amorphous carbon and SiC) or O-rich dust (such as silicates and metal oxides). As we are interested in silicate dust production, we focus this paper only on O-rich stars.

Infrared (IR) spectroscopic observations of O-rich evolved stars show a plethora of features due to molecules and dust \citep{Omont1993, Waters1996, Sylvester1999, Molster2002a}. Molecules, mainly CO, OH, H$_{2}$O, and SiO, dominate the spectra at $\lambda \lesssim 10 \mu$m. At longer wavelengths, dust features become more important. Silicate dust dominates the IR spectra of the majority of dust-producing O-rich AGB stars, and can be found in either amorphous or crystalline form. Amorphous silicates are most abundant, characterised by their broad, smooth features at 10 and 18 $\mu$m. Crystalline silicates exhibit sharp resonance features at 10 $\mu$m and longer wavelengths. The positions and shapes of these are very sensitive to compositional changes, lattice structure, and grain size and morphology, providing a mechanism to identify the chemical composition and mineralogy of the dust grains \citep{Molster2002b, Chihara2002, Koike2003, Min2003, Molster2005}. Comparisons with laboratory data have  primarily identified these grains as Mg-rich olivines and pyroxenes (forsterite: Mg$_{2}$SiO$_{4}$ and enstatite: MgSiO$_{3}$) with little or no iron content \citep{Molster1999, deVries2011}.

The formation of crystalline silicates in the wind of evolved stars is not well understood: crystalline structures are the energetically more favourable atomic arrangement for silicates yet, in most cases, crystalline silicate dust is only a minor component of the circumstellar dust shell. The transition from amorphous to crystalline silicate grains is a thermal process, requiring temperatures in the region of 1040 K \citep{Hallenbeck1998, Fabian2000, Speck2011}. However, there is little consensus on how silicates in stellar outflows gain sufficient energy for this transition to occur. In order to determine the physical conditions under which crystalline silicates form we require a better understanding of the conditions surrounding the star (e.g. wind densities). The formation process of crystalline grains can be constrained by correlating the crystalline fraction with the dust or gas column density.
A correlation with dust density suggests that annealing of amorphous silicate grains heated by radiation is probably the dominant means by which crystals are manufactured \citep{Sogawa1999}, while a correlation with gas density would suggest that direct condensation of crystalline silicates in the wind will dominate \citep[e.g.][]{Tielens1998, GailSed1999}. 

In {\em Infrared Space Observatory} ({\em ISO}) spectra of Milky Way (MW) giant stars, the spectral features due to the crystalline silicates forsterite (Mg$_{2}$SiO$_{4}$) and enstatite (MgSiO$_{3}$) typically only appear around evolved stars if their mass-loss rate is higher than a threshold value of $\sim$10$^{-5}$ M$_\odot$ yr$^{-1}$ \citep{Cami1998, Sylvester1999}. For example, the high-density winds of heavily enshrouded OH/IR stars have measured crystalline fractions of up to $\sim$20 per cent of the silicate mass \citep{Kemper2001, deVries2010}. Conversely, lower mass-loss rate AGB stars, such as Miras, normally lack crystalline silicate dust features in their spectra. This suggests a correlation between lattice order and wind density (e.g. \citealt{Cami1998, Speck2008}). 
Alternatively, contrast effects between amorphous and crystalline silicate grains at different temperatures can mask the characteristic spectral features of the crystalline material (up to the 40 per cent mass fraction level) in infrared spectra of low mass-loss rate AGB stars  \citep{Kemper2001}. If the crystalline and amorphous silicates are in thermal contact with each other, contrast improves, and the detection of smaller amounts ($<$40 per cent) of crystalline silicates becomes possible. This could explain observations of crystalline silicates in lower mass-loss rate objects, such as the recent observations of low-metallicity, low mass-loss rate evolved stars in Galactic globular clusters that show crystalline silicates \citep{Sloan2010,McDonald2011a,Lebzelter2006}.  

\begin{table}
 \begin{minipage}{84mm}
 \caption{LMC and SMC Parameters}
 \label{tab:galProperties}
 \begin{tabular}{@{}lcccc@{}}
   \hline
   \hline
   Parameter    &   LMC    &      Ref.     &      SMC    &    Ref. \\                                                           
\hline             
    Distance,    $d \: (\rm {kpc})     $   &  $ 51 \pm 2    $  &  $   1,2    $ &    $ 61 \pm 2    $    &   $  1,2    $ \\
    Metallicity, $Z \: (Z_{\odot})      $  &  $ 0.5 \pm 0.17 $ &  $   3,4    $ &    $ 0.2 \pm 0.06$    &   $  3,4    $ \\
    Inclination angle  $( ^{\circ} ) $     &  $ 34.7 \pm 6.2 $ &  $   5      $ &    $ 68 \pm 2    $    &   $    6    $ \\
    Gas-to-dust ratio  $ ({\it \Psi}) $    &  $ 200        $   &    $\ldots$   &    $ 500       $      &   $\ldots   $ \\
    E(B-V) (mag)                           &  $ 0.13         $ &  $   7      $ &    $ 0.04        $    &   $   8     $ \\
  \hline
   \multicolumn{5}{p{0.95\textwidth}}{ {\sc References:} (1) \citet{Cioni2000}; (2) \citet{Szewczyk2009}; (3) \citet{Luck1998}; (4) \citet{Meixner2010}; (5) \citet{vanDerMarel2001}; (6)  \citet{Groenewegen2000}; (7) \citet{Massey1995}; (8) \citet{Harris2004}.}
 \end{tabular}
\end{minipage}
\end{table}

Observations of evolved stars in the metal-poor environments of the Magellanic Clouds with the {\em Spitzer Space Telescope} provide an ideal opportunity to explore the occurrence of crystallinity and investigate how the O-rich dust condensation depends on the physical and chemical conditions of the envelope. Within the Large Magellanic Cloud (LMC) and Small Magellanic Cloud (SMC), we have a set of stars with similar characteristics: for instance, they are all at approximately the same distance from the Sun and are assumed to have a single metallicity within each galaxy. This allows us to measure luminosities and dust mass-loss rates from the observed spectral energy distributions. The parameters adopted in our calculations for the Magellanic Cloud stars are listed in Table \ref{tab:galProperties}. The dust-to-gas mass ratios of stars in the Magellanic Clouds are assumed to be lower than those of stars in the Solar neighbourhood and have a linear dependence with metallicity \citep{vanLoon2000,Marshall2004,vanLoon2006}, breaking the degeneracy between the dust column density and the gas density in the outflows of AGB stars. 
By carefully studying the dependence of crystallinity on the dust and gas mass-loss rate, we address the influences of dust density and gas density on the formation of crystalline grains.

This paper is organised as follows: Section \ref{sec:sample} describes the samples. In Section~\ref{sec:analysis} we fit the spectral energy distribution (SED) of each source and measure the strength of crystalline silicate emission bands. Average feature profiles are shown in Section~\ref{sec:results}, and the results are presented. Finally in Section~\ref{sec:discussion} we discuss the implications for the onset of crystallinity and the change in mineralogy with metallicity. The conclusions are summarised in Section~\ref{sec:conclusion}.

\section{The sample} \label{sec:sample}

Our sample contains 69 oxygen-rich AGB stars (O-AGB) and 76 RSG stars in the Magellanic Clouds which were observed spectroscopically with {\it Spitzer}, and 131 Galactic field O-AGBs and RSGs observed with either {\it Spitzer} or {\it ISO}. The {\it Spitzer} spectra cover a wavelength range of 5.2--37.2~$\mu$m while {\it ISO} spectra cover the 2.38--45.2 $\mu$m part of the spectrum. We combine this sample with 39 spectra from 14 Galactic globular clusters to extend the low end of the metallicity range.  

\subsection{LMC sample}

The LMC sample comprises 54 O-AGB and 60 RSG stars observed using the Infrared Spectrograph (IRS, \citealt{Houck2004}) on board \emph{Spitzer}, either as part of the SAGE-Spec legacy program \citep{Kemper2010}, a spectroscopic follow-up to the \emph{Surveying the Agents of a Galaxy's Evolution} project (SAGE-LMC, \citealt{Meixner2006}), or drawn from the homogeneously-reduced archival IRS staring mode spectroscopy within the SAGE-LMC footprint which are incorporated into the SAGE-Spec data delivery. For a full description of the original target selection, observing strategy and the techniques used in the data reduction for the SAGE-Spec legacy program the reader is referred to \cite{Kemper2010}. 

For all sources in the Magellanic Clouds the associated broad-band photometry, including optical \emph{UBVI} photometry from the Magellanic Clouds Photometric Survey \citep{Zaritsky2004}, 2MASS \emph{JHK$_s$} photometry \citep{Skrutskie2006}, mid-IR photometry (IRAC 3.6, 4.5, 5.8, 8.0 and MIPS 24 $\mu $m) and far-IR photometry (MIPS 70 and 160  $\mu $m), was compiled from the {\sc SAGE} catalogue \citep{Meixner2006}. 

The O-AGB and RSG sources in the sample were selected based on the decision-tree classification scheme developed by \cite{Woods2010}. Within the SAGE-Spec catalogue, 197 objects have been classified according to their \emph{Spitzer} IRS spectrum and SED with associated \emph{U, B, V, I, J, H, K$_s$}, IRAC and MIPS photometry. From this we found 40 O-AGB stars, and 19 RSGs. To supplement this LMC sample, we applied the decision-tree classification methodology of \cite{Woods2010} to all the archival IRS observations of point sources within the SAGE-LMC footprint (Woods et al., in prep.), resulting in a further 148 sources. As this work is primarily concerned with crystalline silicates, which become prominent at longer wavelengths ($\lambda > 20$~$\mu$m), for inclusion in the sample we require that the sources were observed with the Long-Low (14.5--37.2 $\mu$m) module on IRS. We limit our study to those spectra which were visually deemed to have a sufficient signal-to-noise for the identification of crystalline silicate features. Table \ref{tab:LMCsample} lists the sources in the LMC sample, along with SAGE-Spec identification (SSID) (see \citealt{Kemper2010}), coordinates, Astronomical Observation Request (AOR), bolometric magnitudes (M$_{{\rm bol}}$), periods and mass-loss rates determined from SED fitting with the {\sc grams} model grid (see Section~\ref{sec:SEDfit}). Only the first entries are shown for guidance, a full version is available in the electronic edition.

\begin{figure}
 \includegraphics[width=94mm]{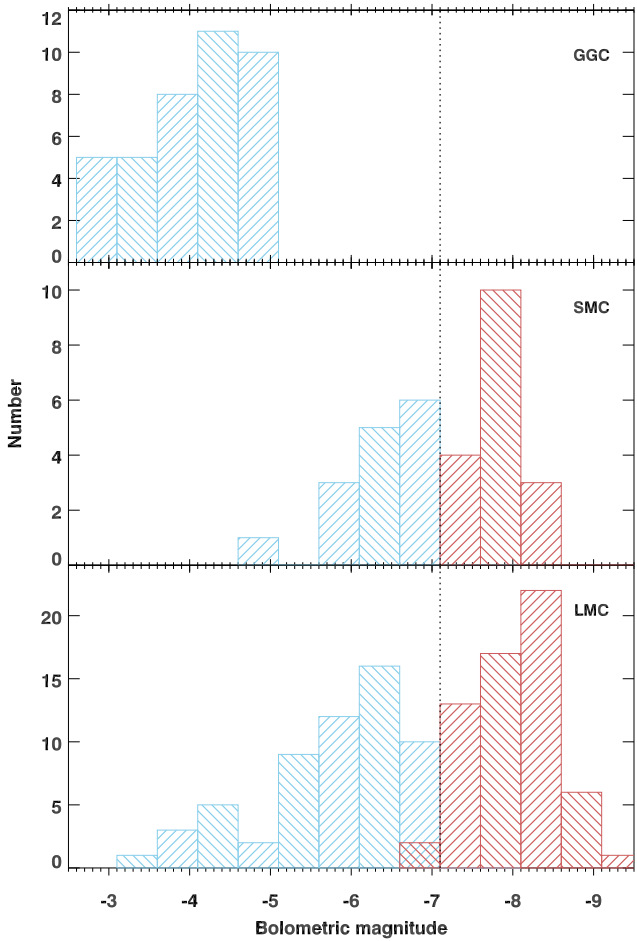}
 \caption{Comparison of the bolometric magnitudes in the LMC, SMC and globular cluster samples. The `classical' luminosity limit for AGB stars is indicated by the dotted line.  Sources to the left of the line are AGB stars (blue) and to the right RSGs (red).}
  \label{fig:Luminosity_function_comparison}
\end{figure}

\begin{table*}
\caption{Details of the sources in the LMC sample for the first few entries. A full version is available in the electronic edition.
For each source the SAGE-Spec identification (SSID), name, coordinates, Astronomical Observation Request (AOR), classification, bolometric magnitudes (M$_{{\rm bol}}$), period and  dust mass-loss rate (${\dot{M}_{dust}}$) is given. }
  \label{tab:LMCsample}
  \begin{tabular}{@{}llccllccc@{}}
   \hline
   \hline
 SSID   &       Name                &	RA	        &	Dec	        &	AOR	        &	Class	& M$_{{\rm bol}}$$^{a}$	&  Period      	& ${\dot{M}_{dust}}$                        \\
        &                           &    (J2000)        &     (J2000)           &                       &               &                       &  (d)          & ($M_{\odot}\ \mathrm{yr}^{-1}$)         \\     
   \hline
4007	&  IRAS 04407$-$7000                 &	04 40 28.49	&	$-$69 55 12.70	&	6077440  	&	O-AGB	&	$-$6.68	&	1179$^{b}$	&	3.61E$-$08                \\
4	&  SSTISAGE J044718.64$-$694220.5    &	04 47 18.63	&	$-$69 42 20.53	&	22400256	&	RSG	&	$-$6.81	&    \ldots       	&	5.01E$-$09                \\
4022	&  MSX LMC 1132                      &	04 49 22.46	&	$-$69 24 34.90	&	10958336  	&	RSG	&	$-$7.97	&	361$^{d}$	&	1.19E$-$08                \\
6	&  SSTISAGE J044934.30$-$690549.2    &	04 49 34.38	&	$-$69 05 49.17	&	24318720 	&	O-AGB	&	$-$5.54	&    \ldots       	&	2.59E$-$09                \\
4024	&  IRAS 04498$-$6842                 &	04 49 41.38	&	$-$68 37 51.00	&	6076672  	&	RSG	&	$-$7.86	&	1280$^{d}$	&	5.43E$-$08                \\
4031	&  IRAS 04509$-$6922                 &	04 50 40.43	&	$-$69 17 31.70	&	6022400  	&	RSG	&	$-$7.51	&	1292$^{d}$	&	2.85E$-$09                \\
8	&  SSTISAGE  J045128.56$-$695550.1   &	04 51 28.58	&	$-$69 55 49.90	&	22400768  	&	O-AGB	&	$-$3.82	&	884$^{b}$	&	2.09E$-$08                \\
4038	&  IRAS 04516$-$6902                 &	04 51 28.94	&	$-$68 57 49.50	&	6020096  	&	O-AGB	&	$-$5.48	&	1090$^{d}$	&	5.12E$-$09                \\
4054	&  2MASS J04524318$-$7047371         &	04 52 43.18	&	$-$70 47 37.80	&	33283584  	&	O-AGB	&	$-$4.22	&    \ldots       	&	7.90E$-$11                \\
\ldots	&  \ldots                   &	\ldots         	&       \ldots	    	&	\ldots   	&	\ldots	&	\ldots  &    \ldots       	&	\ldots                            \\
   \hline																											       
\multicolumn{9}{p{0.99\textwidth}}{(a) Bolometric magnitudes for SSID 1-197 are taken from \cite{Woods2010}, all other values are calculated as described in Sec.~\ref{sec:Mbol}.}\\													       
\multicolumn{9}{p{0.99\textwidth}}{{\sc References:} (b) \cite{Soszynski2008,Soszynski2009}; (c) \cite{Fraser2005,Fraser2008}   ;  (d) \cite{Groenewegen2009} ;  (e) \cite{Pojmanski2002}.}
 \end{tabular}
\end{table*}

\subsection{SMC sample}

The SMC sample of 15 O-AGB and 16 RSG stars was compiled from IRS staring mode observations within the SAGE-SMC \citep{Gordon2011} spatial coverage. Again, these were supplemented with optical, 2MASS, IRAC and MIPS photometry from the {\sc SAGE-SMC} catalogue \citep{Gordon2011}, and the {\sc S$^3$MC} catalogue \citep{Bolatto2007}. Evolved stars were identified from the SAGE-SMC mid-IR classifications of \cite{Boyer2011}, with 57 candidate evolved stars found. IR photometry alone is insufficient to accurately separate the dusty O-AGB and RSG stars from contamination by young stellar objects (YSOs) with strong oxygen-rich dust features, in particular those with a 10-\mum silicate absorption feature. As with the LMC sample, we classify each object according to its \emph{Spitzer} IRS spectrum based on its spectral features, SED, calculated bolometric luminosity, variability information and additional information from the literature, returning 27 O-AGB stars and 18 RSGs. (A full spectral classification for the SMC is ongoing and will be presented in Ruffle et al in prep.).  
Again, sources observed solely with the IRS Short-Low setting or with a poor signal-to-noise ratio were excluded from the sample. A summary of the SMC sample can be found in Table \ref{tab:SMCsample}. 

\begin{table*}
 \caption{Details of the sources in the SMC sample. For each source the name, coordinates, Astronomical Observation Request (AOR), classification, bolometric magnitudes (M$_{{\rm bol}}$), period and dust mass-loss rate (${\dot{M}_{dust}}$) is given.}
 \label{tab:SMCsample}
 \begin{tabular}{@{}l@{\ \ \ \ \ \ \ }c@{\ \ \ \ \ \ \ }c@{\ \ \ \ \ \ \ }l@{\ \ \ \ \ \ \ }l@{\ \ \ \ \ \ \ }c@{\ \ \ \ \ \ \ }c@{\ \ \ \ \ }c@{}}
   \hline
   \hline
    Name	&             RA	&	Dec	&	AOR	&	Class	&	M$_{{\rm bol}}$	&	Period	& 	${\dot{M}_{dust}}$         	 \\
                &            (J2000)    &    (J2000)    &               &               &                       &  (d)          &  ($M_{\odot}\ \mathrm{yr}^{-1}$)       \\
   \hline
 HV 11223	&	00 32 01.90	&	$-$73 22 34.00	&	6019584	        &	O-AGB	&	$-$6.21	&	407$^{a}$	&       1.16E$-$10       \\
 HV 1366	&	00 42 49.90	&	$-$72 55 10.00	&	6017024	        &	O-AGB	&	$-$4.90	&	305$^{a}$	&       3.92E$-$11       \\
 MSX SMC 024	&	00 42 52.23	&	$-$73 50 51.70	&	10663169	&	O-AGB	&	$-$6.64	&	418$^{a}$	&       1.02E$-$08       \\
 MSX SMC 018	&	00 46 31.59	&	$-$73 28 46.50	&	10668800	&	O-AGB	&	$-$6.09	&	915$^{a}$	&       5.46E$-$08       \\
 SMC010889	&	00 48 27.02	&	$-$73 12 12.30	&	25689600	&	RSG	&	$-$8.58	&	\ldots          &       1.75E$-$09       \\
 SMC011709	&	00 48 46.32	&	$-$73 28 20.70	&	25689856	&	RSG	&	$-$7.99	&	\ldots       	&       4.95E$-$10       \\
 PMMR 24	&	00 48 51.79	&	$-$73 22 39.90	&	17409024	&	RSG	&	$-$7.62	&	231$^{b}$	&       2.59E$-$09       \\
 MSX SMC 096	&	00 50 06.40	&	$-$73 28 11.10	&	10667008	&	RSG	&	$-$7.93	&	\ldots       	&       7.86E$-$10       \\
 MSX SMC 055	&	00 50 07.19	&	$-$73 31 25.20	&	10657536	&	RSG	&	$-$7.91	&	1749$^{a}$	&       3.26E$-$08       \\
 MSX SMC 134	&	00 50 44.39	&	$-$72 37 39.00	&	10665216	&	O-AGB	&	$-$5.68	&	248$^{a}$	&       5.96E$-$08       \\
 MSX SMC 109	&	00 51 29.68	&	$-$73 10 44.30	&	10667520	&	RSG	&	$-$8.35	&	\ldots	        &       2.60E$-$09       \\
 HV 11303	&	00 52 09.00	&	$-$71 36 22.00	&	6017536	        &	O-AGB	&	$-$7.07	&	534$^{a}$	&       4.07E$-$10       \\
 BMB$-$B 75	&	00 52 12.87	&	$-$73 08 52.70	&	17409280	&	O-AGB	&	$-$6.74	&	760$^{c}$	&       2.95E$-$08       \\
 HV 11329	&	00 53 39.40	&	$-$72 52 39.20	&	6017280	        &	O-AGB	&	$-$6.40	&	377$^{a}$	&       3.27E$-$10       \\
 MSX SMC 000	&	00 55 18.10	&	$-$72 05 31.00	&	10666240	&	RSG	&	$-$7.10	&	544$^{a}$	&       2.77E$-$10       \\
 MSX SMC 168	&	00 55 26.76	&	$-$72 35 56.10	&	10668288	&	RSG	&	$-$8.00	&	\ldots       	&       1.76E$-$09       \\
 HV 838	        &	00 55 38.30	&	$-$73 11 42.00	&	6017792	        &	RSG	&	$-$7.21	&	654$^{a}$	&       9.49E$-$11       \\
 HV 11366	&	00 56 54.80	&	$-$72 14 07.00	&	6018304	        &	O-AGB	&	$-$6.27	&	183$^{a}$	&       4.23E$-$11       \\
 HV 12149	&	00 58 50.30	&	$-$72 18 34.00	&	6019328	        &	O-AGB	&	$-$7.04	&	745$^{a}$	&       1.88E$-$09       \\
 SMC046662	&	00 59 35.04	&	$-$72 04 06.20	&	25690112	&	RSG	&	$-$8.07	&	321$^{b}$	&       3.10E$-$09       \\
 MSX SMC 181	&	01 00 48.18	&	$-$72 51 02.10	&	10665728	&	RSG	&	$-$7.28	&	1062$^{a}$	&       8.85E$-$09       \\
 HV 11423	&	01 00 54.80	&	$-$71 37 55.00	&	25691136	&	RSG	&	$-$7.78	&	\ldots       	&       2.94E$-$09       \\
 SMC052334	&	01 01 54.16	&	$-$71 52 18.80	&	25690368	&	RSG	&	$-$7.68	&	\ldots       	&       3.42E$-$10       \\
 SMC055188	&	01 03 02.38	&	$-$72 01 52.90	&	27603200	&	RSG	&	$-$7.54	&	530$^{c}$	&       3.91E$-$09       \\
 SMC55681	&	01 03 12.98	&	$-$72 09 26.50	&	25687808	&	RSG	&	$-$7.79	&	\ldots       	&       2.96E$-$09       \\
 MSX SMC 234	&	01 03 42.35	&	$-$72 13 42.80	&	27522048	&	O-AGB	&	$-$6.49	&	\ldots       	&       4.05E$-$08       \\
 HV 1963	&	01 04 26.80	&	$-$72 34 40.00	&	6018048	        &	O-AGB	&	$-$6.74	&	330$^{a}$	&       7.28E$-$11       \\
 IRAS F01066$-$7332 &	01 08 10.27	&	$-$73 15 52.30	&	17409536	&	O-AGB	&	$-$5.64	&	882$^{c}$	&       3.35E$-$08       \\
 HV 12956	&	01 09 02.25	&	$-$71 24 10.20	&	27528960	&	O-AGB	&	$-$6.27	&	523$^{c}$	&       5.79E$-$08       \\
 MSX SMC 149	&	01 09 38.24	&	$-$73 20 02.40	&	10668032	&	RSG	&	$-$7.98	&	\ldots       	&       5.18E$-$09       \\
 SMC083593	&	01 30 33.92	&	$-$73 18 41.90	&	25690880	&	O-AGB	&	$-$6.80	&	\ldots       	&       4.11E$-$09       \\
\hline																			
\multicolumn{8}{p{0.9\textwidth}}{{\sc References:} (a) \cite{Groenewegen2009} ; (b) \cite{Pojmanski2002} ;  (c) \cite{Soszynski2011}.}
 \end{tabular}
\end{table*}

\subsection{Galactic field sample}

To provide a solar metallicity comparison, our sample also includes \emph{ISO} Short Wavelength Spectrometer (SWS; \citealt{DeGraauw1996}) spectra of 79 O-AGB, 9 probable O-AGBs and 22 RSG stars in the Milky Way, covering a wavelength range of 2.38 -- 45.2 $\mu$m. This Galactic field sample was compiled from the ISO/SWS database\footnote{\url{http://isc.astro.cornell.edu/~sloan/library/swsatlas/aot1.html}}, which contains 1271 observations in full-scan mode. This database contains observations obtained for a wide range of individual observing projects, which have been uniformly reduced by \cite{Sloan200archive} and classified based on the overall shape of the SED and their IR spectral features (such as oxygen- or carbon-rich dust emission, etc.) by \cite{Kraemer2002}.
From this database, we selected those sources with optically-thin, oxygen-rich dust emission (group 2.SE) sources dominated by emission from warm silicate dust (groups 3.SE and 3.SB) and the optically-thick sources with cool silicate dust (groups 4.SA, 4.SB, 4.SC and 4.SE) which have a sufficient signal-to-noise ratio to distinguish the solid state features. Sources which have no significant dust emission (only SiO absorption at 8 $\mu$m), incomplete spectra and those which did not fulfil the criteria to be classed as an O-AGB or RSG by the \cite{Woods2010} decision-tree were eliminated from the sample. We also require the spectra to show the amorphous silicate features in order to fit their spectral energy distribution using the {\sc grams} model grid. We verified in the literature that the selected sources were classified as either an O-AGB or RSG.
The resulting list of 110 spectra is shown in Table \ref{tab:GalaticSample}. In instances where a source was observed by the SWS on multiple occasions, we analyse each spectrum independently. The TDT (Target Dedicated Time) number is used to identify the particular spectra for each source.
Since both RSGs and O-AGB stars are red, luminous and exhibit similar dust emission characteristics, confusion exists between these two types of sources. Sources for which reliable estimates for the distance are available or are known OH masers are further sub-divided into O-AGB and RSG stars.  We consider distances from the revised {\it Hipparcos} catalogue \citep{Hipparcos} to be reliable if the relative error is less than 20 per cent. The variability types have been taken from the Combined General Catalogue of Variable stars (GCVS; \citealt{Kholopov1998, Samus2004}). 

\begin{table*}
 \caption{Details of the sources in the Galactic \emph{ISO} SWS sample. A full version is available in the electronic edition. For each source the name, coordinates, TDT (Target Dedicated Time) number, classification, distance, period and dust mass-loss rate (${\dot{M}_{dust}}$) is given.}
 \label{tab:GalaticSample}
 \begin{tabular}{@{}lcrllcccc@{}}
   \hline
   \hline
   Name	  &	RA	  &\multicolumn{1}{|c|}{Dec}	  &   \multicolumn{1}{|c|}{TDT}     & 	Class	  &	Distance	  &  Ref.	  &	Period$^{a}$	  &	${\dot{M}_{dust}}$ \\
          &  (J2000)      &\multicolumn{1}{|c|}{(J2000)}  &           &	          &    (pc)  	    	  &  distance	  &	 (d)  	          &	 ($M_{\odot}\ \mathrm{yr}^{-1}$)         \\
   \hline
	S Scl	&	00 15 22.18	&	$-$32 02 43.4	& 37102018        &	O-AGB	&	430	&	16	&	363	&       1.35E$-$09      \\
	S Scl	&	00 15 22.18	&	$-$32 02 43.4	& 73500129        &	O-AGB	&	430	&	16	&	363	&       1.13E$-$09      \\
	WX Psc	&	01 06 25.96	&	 12 35 53.1	& 39502217        &	O-AGB	&	833	&	4	&	660	&       1.92E$-$07      \\
	WX Psc	&	01 06 25.96	&	 12 35 53.1	& 76101413        &	O-AGB	&	833	&	4	&	660	&       1.92E$-$07      \\
 OH 127.8 $+$0.0 &	01 33 50.60	&	 62 26 47.0	& 44301870        &	O-AGB	&	\ldots	&	\ldots	&	\ldots	&       2.36E$-$06      \\
 OH 127.8 $+$0.0 &	01 33 50.60	&	 62 26 47.0	& 78800604        &	O-AGB	&	\ldots	&	\ldots	&	\ldots	&       6.25E$-$07      \\
	SV Psc	&	01 46 35.30	&	 19 05 04.0	& 80501620        &	O-AGB	&	380	&	14	&	102	&       3.08E$-$09      \\
	Mira	&	02 19 20.78	&	$-$02 58 36.2	& 45101201        &	O-AGB	&	128	&	8 	&	332	&       1.18E$-$08      \\
	SU Per	&	02 22 06.93	&	 56 36 15.1	& 43306303        &	RSG	&	1900	&	10	&	533	&       1.22E$-$08      \\
	\ldots  &       \ldots	    	&\multicolumn{1}{|c|}{\ldots} & \multicolumn{1}{|c|}{\ldots}	       & \ldots	&	\ldots  &    \ldots	&	\ldots	&	\ldots           	\\
   \hline
\multicolumn{9}{p{0.9\textwidth}}{(a) Periods are taken from the GCVS \citep{Kholopov1998, Samus2004}.}\\
\multicolumn{9}{p{0.9\textwidth}}{{\sc References:} (1) \cite{Hipparcos}; (2) \cite{Chen2008}; (3) \cite{deBeck2010}; (4) \cite{Decin2007}; (5) \cite{Herman1986}; (6) \cite{Humphreys1978}; (7) \cite{Justtanont2006}; (8) \cite{Knapp1988};  (9) \cite{LeSidaner1996}; (10) \cite{Levesque2005}; (11) \cite{Loup1993}; (12) \cite{Massey1991}; (13) \cite{Olivier2001}; (14) \cite{Olofsson2002};  (15) \cite{vanLangevelde1990}; (16) \cite{Whitelock2008}; (17) \cite{Xiong1994}; (18) \cite{Young1995}.}
 \end{tabular}
\end{table*}

\subsubsection{Spitzer P1094 Galactic sample}

\emph{Spitzer} low-resolution spectra of 21 faint Galactic O-AGB stars were observed as part of Program 1094. Spectral data reduction was done in {\sc smart} \citep{Higdon2004} in a manner similar to that summarised by \cite{Furlan2006} and \cite{Sargent2009}. Many of these sources are not well known in the literature and ancillary data including photometry, periods, variability type and distance indicators were difficult to find in the literature. As before we classify these sources using the decision tree, however, these classifications may be less robust than the other samples due to uncertainty in the SED shape. This means our sample may suffer some contamination by oxygen-rich post-AGB sources. The Galactic \emph{Spitzer} sources are listed in Table~\ref{tab:P1094Sample}.

\begin{table*}
 \caption{Details of the sources in the Galactic \emph{Spitzer} IRS sample. For each source the name, coordinates, Astronomical Observation Request (AOR), classification, period and dust mass-loss rate (${\dot{M}_{dust}}$) is given.}
 \label{tab:P1094Sample}
 \begin{tabular}{@{}l@{\ \ \ \ \ \ \ \ \ \ \ \ }c@{\ \ \ \ \ \ \ \ \ }r@{\ \ \ \ \ \ \ }c@{\ \ \ \ \ \ \ \ \ \ \ \ }c@{\ \ \ \ \ \ \ \ \ }c@{\ \ \ \ \ \ \ \ \ }c@{}}
   \hline
   \hline
Name	                &	RA	        &\multicolumn{1}{|c|}{Dec}       &	AOR	&	Class$^{a}$	& Period$^{b}$	&    ${\dot{M}_{dust}}$      \\
      	                &      (J2000)     	&\multicolumn{1}{|c|}{(J2000)}   &          	&	     	&     (d)      	&  ($M_{\odot}\ \mathrm{yr}^{-1}$)   \\
   \hline
IRAS 00534$+$6031	&	00 56 28.41	&	60 47 09.25	&	6082816	&	O-AGB	&	410.0	&       5.44E$-$09       \\
IRAS 00589$+$5743	&	01 01 58.43	&	57 59 47.86	&	6083072	&	\ldots	&	188.0	&       5.07E$-$08       \\
IRAS 03434$+$5818	&	03 47 31.26	&	58 28 10.41	&	6082304	&	\ldots	&	\ldots	&       1.14E$-$07       \\
IRAS 05314$+$2020 	&	05 34 26.48	&	20 22 53.78	&	6083584	&	\ldots	&	240.0	&       2.99E$-$09       \\
IRAS 08425$-$5116	&	08 44 04.77	&	 $-$51 27 43.74	&	6078976	&	O-AGB	&	\ldots	&       4.65E$-$07       \\
IRAS 13581$-$5444	&	14 01 28.90	&	 $-$54 59 00.08	&	6081536	&	O-AGB	&	\ldots	&       3.05E$-$07       \\
IRAS 16523$+$0745	&	16 54 46.42	&	07 40 27.07	&	6083328	&	\ldots	&	274.0	&       5.71E$-$09       \\
IRAS 17030$-$3053	&	17 06 14.07	&	 $-$30 57 38.83	&	6084864	&	O-AGB	&	\ldots	&       6.58E$-$07       \\
IRAS 17276$-$2846	&	17 30 48.30	&	 $-$28 49 02.39	&	6080768	&	\ldots	&	\ldots	&       7.06E$-$07       \\
IRAS 17304$-$1933	&	17 33 22.13	&	 $-$19 35 51.75	&	6084352	&	O-AGB	&	\ldots	&       1.16E$-$07       \\
IRAS 17338$-$2140	&	17 36 52.23	&	 $-$21 42 41.08	&	6085632	&	O-AGB	&	\ldots	&       1.89E$-$07       \\
IRAS 17347$-$2319	&	17 37 46.29	&	 $-$23 20 53.48	&	6080000	&	O-AGB	&	\ldots	&       4.48E$-$07       \\
IRAS 17413$-$3531	&	17 44 43.45	&	 $-$35 32 35.28	&	6081280	&	\ldots	&	\ldots	&       1.91E$-$07       \\
IRAS 17513$-$3554	&	17 54 42.02	&	 $-$35 54 51.38	&	6080256	&	O-AGB	&	\ldots	&       3.68E$-$07       \\
IRAS 18195$-$2804	&	18 22 40.17	&	 $-$28 03 07.61	&	6079744	&	O-AGB	&	\ldots	&       4.16E$-$07       \\
IRAS 18231$+$0855	&	18 25 33.36	&	08 56 46.62	&	6082560	&	\ldots	&	\ldots	&       1.76E$-$08       \\
IRAS 18279$-$2707	&	18 31 03.68	&	 $-$27 05 40.93	&	6085376	&	O-AGB	&	\ldots	&       2.97E$-$08       \\
IRAS 18291$-$2900 	&	18 32 19.55	&	 $-$28 58 09.81	&	6084096	&	\ldots	&	\ldots	&       4.17E$-$08       \\
IRAS 19256$+$0254	&	19 28 08.39	&	03 00 25.32	&	6081792	&	O-AGB	&	\ldots	&       1.14E$-$07       \\
IRAS 19456$+$1927	&	19 47 49.65	&	19 35 22.67	&	6081024	&	O-AGB	&	\ldots	&       1.92E$-$07       \\
GX Tel	                &	20 11 48.22	&	 $-$55 25 26.65	&	6082048	&	O-AGB	&	340.8	&       1.63E$-$08       \\
   \hline
\multicolumn{7}{p{0.9\textwidth}}{(a) The available data were insufficient for determining whether some sources were an O-AGB or RSG; these sources are not assigned a classification.}\\
\multicolumn{7}{p{0.9\textwidth}}{(b) Periods are taken from the GCVS \citep{Kholopov1998, Samus2004}.}\\
 \end{tabular}
\end{table*}

\subsection{Globular cluster sample}

To complement these sources, \emph{Spitzer} IRS observations of oxygen-rich sources in 14 Galactic globular clusters (GGC) allow the metallicity range we investigate to be extended to lower metallicities. The globular cluster sample consists of 39 O-AGB stars (observed in the Short- and Long-Low) which are discussed by \cite{Sloan2010} and \cite{McDonald2011}. Both these studies compute bolometric luminosities, periods and dust mass-loss rates for each object allowing a ready comparison to the Magellanic Cloud stars, however, it should be noted that the mass-loss rates are obtained via an alternative method to that discussed in Section~\ref{sec:SEDfit}. For consistency we recalculate the mass-loss rates using the {\sc grams} models to obtain a uniform treatment of the mass-loss rates and directly compare the globular cluster sources to the other samples. Table~\ref{tab:GlobClusterSample} lists the sources considered in this work.

\begin{table*}
 \caption{Details of the sources in the globular cluster sample. Distances, [Fe/H] values, periods and bolometric magnitudes for NGC 5139 ($\omega$ Cen) were taken from \protect\cite{McDonald2011}; values for these parameters for all other clusters are from \protect\cite{Sloan2010}.}
 \label{tab:GlobClusterSample}
 \begin{tabular}{@{}l@{\ \ \ }l@{\ \ \ \ \ }c@{\ \ \ \ }c@{\ \ \ \ \ }c@{\ \ \ \ }c@{\ \ \ \ }c@{\ \ \ \ }c@{\ \ \ \ }c@{\ \ \ \ }c@{\ \ \ \ }c@{}}
   \hline
   \hline
Cluster       	& Star name	&	RA	        &	Dec	        &	AOR	      & Distance       	&	[Fe/H]       	& Period       	&  M$_{\rm bol}$       	&${\dot{M}_{dust}}$	\\	
           	&		&	(J2000)	        &	(J2000)	        &		        & (kpc)    	&	        	& (d)         	&       	& ($M_{\odot}\ \mathrm{yr}^{-1}$)       \\	
   \hline
NGC 362	        &	V16	&	01 03 15.10	&	$-$70 50 32.3	&	21740800	&	9.25	&	$-$1.20	&	138	&	$-$4.10	&       1.84E$-$09           \\      
NGC 362	        &	V2	&	01 03 21.85	&	$-$70 54 20.1	&	21740800	&	9.25	&	$-$1.20	&	89	&	$-$3.50	&       5.80E$-$10           \\      
NGC 5139	& Leid 61015	&	13 25 21.33	&	$-$47 36 53.9	&	27854592	&	5.30	&	$-$1.71	&	\ldots	&	$-$3.29	&       7.92E$-$11           \\      
NGC 5139	& Leid 42044	&	13 26 05.35	&	$-$47 28 20.6	&	27854848	&	5.30	&	$-$1.37	&	0.3	&	$-$3.01	&       1.30E$-$10           \\      
NGC 5139	& Leid 33062	&	13 26 30.19	&	$-$47 24 27.8 	&	27856128	&	5.30	&	$-$1.08	&	110	&	$-$3.61	&       4.36E$-$09           \\      
NGC 5139	& Leid 44262	&	13 26 46.36	&	$-$47 29 30.4 	&	27852800	&	5.30	&	$-$0.80	&	149.4	&	$-$3.27	&       6.25E$-$10           \\      
NGC 5139	& Leid 44262	&	13 26 46.36	&	$-$47 29 30.4	&	21741056	&	5.30	&	$-$1.63	&	149	&	$-$3.88	&       1.30E$-$09           \\      
NGC 5139	& Leid 44277	&	13 26 47.72	&	$-$47 29 29.0 	&	27855104	&	5.30	&	$-$1.37	&	124	&	$-$3.19	&       2.40E$-$10           \\      
NGC 5139	& Leid 56087	&	13 26 48.05	&	$-$47 34 56.8	&	27853824	&	5.30	&	$-$1.84	&	\ldots	&	$-$3.08	&       1.38E$-$10           \\      
NGC 5139	& Leid 43351	&	13 26 55.02	&	$-$47 28 45.9	&	27854336	&	5.30	&	$-$0.98	&	\ldots	&	$-$2.79	&       7.39E$-$11           \\      
NGC 5139	& Leid 55114	&	13 26 55.55	&	$-$47 34 23.4	&	27854080	&	5.30	&	$-$1.45	&	\ldots	&	$-$3.16	&       1.59E$-$10           \\      
NGC 5139	& Leid 41455	&	13 27 15.82	&	$-$47 27 54.6	&	27855360	&	5.30	&	$-$1.22	&	90	&	$-$3.05	&       1.64E$-$10           \\      
NGC 5139	& Leid 35250	&	13 27 37.73 	&	$-$47 25 17.5	&	27855872	&	5.30	&	$-$1.06	&	65	&	$-$3.05	&       4.28E$-$10           \\      
NGC 5927	&	V3	&	15 28 00.13	&	$-$50 40 24.6	&	21741824	&	7.76	&	$-$0.35	&	297	&	$-$4.64	&       1.10E$-$08           \\      
NGC 5927	&	V1	&	15 28 15.17	&	$-$50 38 09.3	&	21741568	&	7.76	&	$-$0.35	&	202	&	$-$4.13	&       6.21E$-$09           \\      
NGC 6356	&	V5	&	17 23 17.06	&	$-$17 46 24.5	&	21743104	&	15.35	&	$-$0.50	&	220	&	$-$4.61	&       1.47E$-$09           \\      
NGC 6356	&	V3	&	17 23 33.30	&	$-$17 48 07.4	&	21743104	&	15.35	&	$-$0.50	&	223	&	$-$4.06	&       1.19E$-$09           \\      
NGC 6356	&	V1	&	17 23 33.72	&	$-$17 49 14.8	&	21743104	&	15.35	&	$-$0.50	&	227	&	$-$4.54	&       1.33E$-$08           \\      
NGC 6356	&	V4	&	17 23 48.00	&	$-$17 48 04.5	&	21743104	&	15.35	&	$-$0.50	&	211	&	$-$4.42	&       9.85E$-$10           \\      
NGC 6352	&	V5	&	17 25 37.52	&	$-$48 22 10.0	&	21742848	&	5.89	&	$-$0.69	&	177	&	$-$4.27	&       3.20E$-$09           \\      
NGC 6388	&	V4	&	17 35 58.94	&	$-$44 43 39.8	&	21743360	&	12.42	&	$-$0.57	&	253	&	$-$4.68	&       6.21E$-$09           \\      
NGC 6388	&	V3	&	17 36 15.04	&	$-$44 43 32.5	&	21743360	&	12.42	&	$-$0.57	&	156	&	$-$3.83	&       3.55E$-$10           \\      
Terzan 5        &	V7	&	17 47 54.33	&	$-$24 49 54.6	&	21744128	&	6.64	&	$-$0.08	&	377	&	$-$4.98	&       1.88E$-$08           \\      
Terzan 5        &	V2	&	17 47 59.46	&	$-$24 47 17.6	&	21743872	&	6.64	&	$-$0.08	&	217	&	$-$4.41	&       4.87E$-$09           \\      
Terzan 5        &	V5	&	17 48 03.40	&	$-$24 46 42.0	&	21744128	&	6.64	&	$-$0.08	&	464	&	$-$5.09	&       2.27E$-$08           \\      
Terzan 5        &	V8	&	17 48 07.18	&	$-$24 46 26.6	&	21744128	&	6.64	&	$-$0.08	&	261	&	$-$5.04	&       2.68E$-$09           \\      
Terzan 5        &	V6	&	17 48 09.25	&	$-$24 47 06.3	&	21744128	&	6.64	&	$-$0.08	&	269	&	$-$4.72	&       2.31E$-$08           \\      
NGC 6441	&	V2	&	17 50 16.16	&	$-$37 02 40.5	&	21744384	&	13.00	&	$-$0.56	&	145	&	$-$3.98	&       1.31E$-$09           \\      
NGC 6441	&	V1	&	17 50 17.09	&	$-$37 03 49.7	&	21744384	&	13.00	&	$-$0.56	&	200	&	$-$4.26	&       1.38E$-$09           \\      
IC 1276	        &	V3	&	18 10 50.79	&	$-$07 13 49.1	&	21742080	&	5.40	&	$-$0.69	&	300	&	$-$4.72	&       1.24E$-$08           \\      
IC 1276	        &	V1	&	18 10 51.55	&	$-$07 10 54.5	&	21742080	&	5.40	&	$-$0.69	&	222	&	$-$4.60	&       1.57E$-$09           \\      
NGC 6637	&	V4	&	18 31 21.88	&	$-$32 22 27.7	&	21745664	&	8.95	&	$-$0.66	&	200	&	$-$4.44	&       1.04E$-$08           \\      
NGC 6637	&	V5	&	18 31 23.44	&	$-$32 20 49.5	&	21745664	&	8.95	&	$-$0.66	&	198	&	$-$4.44	&       2.22E$-$09           \\      
NGC 6712	&	V7	&	18 52 55.38	&	$-$08 42 32.5	&	21745920	&	7.05	&	$-$0.97	&	193	&	$-$4.34	&       2.01E$-$09           \\      
NGC 6712	&	V2	&	18 53 08.78	&	$-$08 41 56.6	&	21745920	&	7.05	&	$-$0.97	&	109	&	$-$3.91	&       2.59E$-$09           \\      
NGC 6760	&	V3	&	19 11 14.31	&	01 01 46.6	&	21746176	&	8.28	&	$-$0.48	&	251	&	$-$4.71	&       6.50E$-$09           \\      
NGC 6760	&	V4	&	19 11 15.03	&	01 02 36.8	&	21746432	&	8.28	&	$-$0.48	&	226	&	$-$4.34	&       6.40E$-$09           \\      
Palomar 10	&	V2	&	19 17 51.48	&	18 34 12.7	&	21746944	&	5.92	&	$-$0.10	&	393	&	$-$4.96	&       8.85E$-$09           \\      
NGC 6838	&	V1	&	19 53 56.10	&	18 47 16.8	&	21746944	&	3.96	&	$-$0.73	&	179	&	$-$4.00	&       2.32E$-$09           \\  
   \hline																									
\end{tabular}			
\end{table*}

\section{Analysis} \label{sec:analysis}

\subsection{Bolometric magnitudes} \label{sec:Mbol}

The bolometric magnitudes of our sources in the LMC and SMC samples are listed in Tables~\ref{tab:LMCsample} and \ref{tab:SMCsample}.  These were calculated via a simple trapezoidal integration of the IRS spectrum and optical/infrared photometry, to which a Wien tail was fitted in the optical, while a Rayleigh-Jeans tail was fitted to the long-wavelength data.  
For sources with little infrared excess, bolometric magnitudes were also calculated using the SED-fitting code described by \cite{McDonald2009}.   This code performs a $\chi^2$-minimisation between the observed SED (corrected for interstellar reddening) and a grid of {\sc bt-settl} stellar atmosphere models \citep{Allard2011} which are scaled in flux to derive a bolometric luminosity.  This provided a better fit to the optical and near-IR photometry, than fitting a Planck function. For the most enshrouded stars, fitting the SED with `naked' stellar photosphere models leads to a underestimation of the temperature and luminosity due to circumstellar reddening and hence the integration method is preferred for very dusty sources. The distances and \emph{E(B-V)} values adopted in our calculations are listed in Table~\ref{tab:galProperties}. 

The `classical' upper luminosity limit for AGB stars, based on the core-mass--luminosity relationship, is M$_{bol}$ = 7.1 \citep{Wood1983}; this is not an absolute limit, with a few AGB sources occasionally traversing this limit during thermal pulses \citep{Wood1992}. We use this value as a discriminant between the O-AGB and RSGs in our sample, unless there is good evidence in the literature to support a different classification (for instance optical spectra with lithium absorption or OH maser emission would suggest the object is an O-AGB star). Fig.~\ref{fig:Luminosity_function_comparison} compares the bolometric magnitudes of the O-AGB and RSG sources in the globular cluster, LMC and SMC samples. In the LMC we probe the full AGB luminosity range, however, in the SMC the observed sample is biased to the brightest AGB sources and supergiants.

\subsection{Determining the gas and dust mass-loss rates} \label{sec:SEDfit}

We derive individual dust mass-loss rates (${\dot{M}_{dust}}$) of O-AGB and RSG stars by fitting their infrared SED with the oxygen-rich subset of the `Grid of Red supergiant and Asymptotic giant branch ModelS' ({\sc grams}; \citealt{Sargent2011}). This approach avoids the onerous task of determining these quantities by individually modelling via radiative transfer (RT) the nuances of the IRS and SWS spectra. The coverage of the  oxygen-rich {\sc grams} models has been compared to the observed colours and magnitudes of the SAGE-Spec sample and was found to be in good agreement (see figs.~2--7 from \citealt{Sargent2011}). 
 Gas mass-loss rates can be obtained by scaling the dust mass-loss rates using an assumed gas-to-dust-ratio ($\Psi$) based on the metallicity. Table \ref{tab:g2d} lists the values of $\Psi$ adopted in our calculations. We assume that the field stars in the Milky Way come from a single solar metallicity population with $\Psi_{\odot} = 100$ \citep{Knapp1993}. For the globular clusters, the gas-to-dust-ratio was determined from the metallicity listed in Table \ref{tab:GlobClusterSample} using the following relation: $\Psi = 100 \times 10^{-\mathrm{[Fe/H]}}$. 
However, it should be noted that the gas-to-dust-ratio metallicity dependence is only weakly constrained and thus the adopted gas-to-dust ratios are highly uncertain.


\begin{table}
 \caption{Summary of gas-to-dust-ratios.}
 \label{tab:g2d} 
 \centering
 \begin{tabular}{@{}lc@{}}
   \hline
   \hline
Environment & Gas-to-dust ratio  $ ({\it \Psi}) $ \\	
			
\hline			
MW	          &	100	\\
LMC	          &	200	\\
SMC	          &	500	\\
NGC 362	          &	1585	\\
NGC 5139	  &	4266	\\
NGC 5927	  &	224	\\
NGC 6356	  &	316	\\
NGC 6352	  &	490	\\
NGC 6388	  &	372	\\
Terzan 5      	  &	120	\\
NGC 6441	  &	363	\\
IC 1276	          &	490	\\
NGC 6637	  &	457	\\
NGC 6712	  &	933	\\
NGC 6760	  &	302	\\
Palomar 10	  &	126	\\
NGC 6838	  &	537	\\
\hline			
 \end{tabular}
\end{table}


The optical/infrared broad-band photometry for all sources in the Magellanic Clouds, were combined with `synthetic' \emph{IRAS} 12-$\mu $m fluxes obtained by convolving the spectrum with the 12-$\mu $m \emph{IRAS} band profile\footnote{The \emph{IRAS} filter bandpasses and zero points can be found in \cite{Neugebauer1984}.}.	
In instances where there was more than one epoch of data for a band, the photometry was averaged to minimise variability effects (due to pulsations) in the fitting. Additionally, if a source was missing photometric data, the IRS spectrum was convolved with the appropriate broadband photometric filter and included in the SED, to better constrain the fit parameters and prevent the fit being limited by a lack of data points in a given region. 

The SEDs for the Milky Way sources were constructed using 2MASS \emph{JHK$_s$} photometry combined with {\em IRAS} 12, 60 and 100 $\mu $m data from the Point Source Catalogue \citep{Beichman1988}, and the Faint Source Catalogue \citep{Moshir1989}. To obtain a uniform data set we supplemented the SED with convolved photometry from the \emph{ISO} spectra in the IRAC and MIPS 24 $\mu $m bands to obtain synthetic fluxes. These were calculated using the IDL routine {\sc spitzer\_synthphot}, which is available from the Spitzer Science Center.
Synthetic photometry was also derived for desired photometric bands from the O-rich {\sc grams} models according to the convolution method outlined above\footnote{In the case of 2MASS relative spectral response curves are provided by \cite{Cohen2003}.}.

Variable AGB stars can show large absolute magnitudes amplitudes in the optical, typically in the range of $ \Delta M_{V}  \lesssim  11 $ mag \citep{Samus2004}. Infrared amplitudes are less than this with \emph{K}-band amplitudes usually $\Delta M_{K} \lesssim 2.3$ mag and IRAC 3.6 $\mu $m $ \Delta M_{3.6} \lesssim 2.0 $ mag \citep{McQuinn2007,Vijh2009}. Due to the large variability in optical \emph{UBVI} data we chose to exclude these points from the SED fitting. 

Long-period Mira variables can also show significant variation in amplitude due to pulsations in the mid-IR: multi-epoch \emph{ISO}-SWS spectral studies of the variable stars T Cep, R Aql and R Cas covering minimum to post-maximum pulsation phases over a single pulsation cycle allow us to characterise the variations in amplitude at longer wavelengths ($\lambda > 8.0 \, \mum$). Synthetic photometry was obtained from the spectra at each epoch for the \emph{IRAS} [12] and MIPS [24] bands, and the difference in magnitude between photometric maximum and minimum was used as a measure of the amplitude. We found that $ \Delta M_{12} \lesssim 0.4 $ mag and $ \Delta M_{24} \lesssim 0.6 $. For all sources the photometric uncertainties in these bands were modified to account for this change in magnitude during a pulsation cycle. At near-IR wavelengths we also account for changes in magnitude due to pulsations by adjusting the photometric uncertainties to reflect the differences in magnitude across a pulsation cycle. 

We do not correct the SEDs for interstellar reddening. Due to the close proximity ($d  \lesssim 4$ kpc) of the Galactic sources and the low column densities along each line-of-sight to the Magellanic Clouds the dereddening correction is negligible compared to the uncertainty in the photometric measurements if variability is considered. 

A \chisq--minimisation analysis was used to fit the convolved model fluxes to the observed SEDs. As the grid is calculated for sources at $50$ kpc, we scale the convolved model fluxes to match those of the star before commencing the \chisq fitting. The scaling factor was calculated such that it minimised the difference between the source and the model, this allows the distance $d$ to be a free parameter. 

The reduced \chisq, weighted by photometric uncertainties is given by: 

\begin{equation}
  \chi^2= \frac{1}{\rm {N}}\sum_{i=1}^{N} \left(\frac{ (\mathrm{log}_{10}[ \lambda F_{\lambda}(i)] - \mathrm{log}_{10} [\lambda M_{\lambda}(i)])^2}{\sigma (\mathrm{log}_{10}[\lambda F_{\lambda}(i)])^2}\right),
\end{equation}
where $ \lambda F_{\lambda}$ is the observed broadband photometry with errors,  
\begin{equation}
\sigma (\mathrm{log}_{10}[\lambda F_{\lambda}]) = \frac{1}{\mathrm{ln}10} \frac{\sigma( \lambda F_{\lambda})}{\lambda F_{\lambda}}, 
\end{equation}
and $ \lambda M_{\lambda}$ is the scaled model fluxes for $N$ measurements. 
This fitting process is repeated for each model SED with a scaling factor and \chisq ~value returned for each model. In order to investigate the onset of crystallinity determining accurate dust mass-loss rates is vital. By fitting in the $\mathrm{log}_{10}[ \lambda F_{\lambda}]$ plane we are effectively gives more weight to the longer wavelength bands where the dust grains emit most of their flux and obtain a better fit to the shape of the SED. 
This method relies upon the source being well sampled across a range of photometric bands, however there are some sources where there is no 2MASS {\em JHKs}, IRAS [60] or MIPS [70], so the overall fit is more weakly constrained.

\begin{table}
 \begin{minipage}{84mm}
 \caption{The scale factor range.}
 \label{tab:scale}
 \centering
 \begin{tabular}{@{}lcc}
   \hline
   \hline
    Galaxy &   Distance (kpc) &  Scale factor  \\
   \hline
           LMC    &   $51 \pm 3  $   & $0.86 - 1.09$   \\
           SMC    &   $61 \pm 3  $   & $0.61 - 0.74$   \\
        Milky Way &   $ < 4      $   &  $  > 156   $    \\
   \hline
 \end{tabular}
 \end{minipage}
\end{table}


\begin{figure}
\includegraphics[width=84mm]{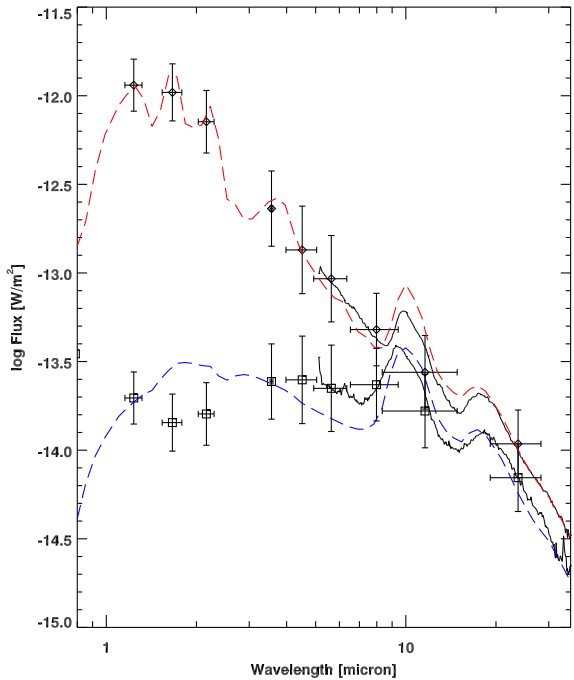}
 \caption{Best-fitting models to the broadband photometry for a RSG (MSX SMC 234; dashed red) and an O-AGB star (MSX SMC 149; dashed blue) in the SMC. The solid line is the observed spectra for each star. Similarly the square and diamond points are the respective observed photometry for the O-AGB and RSG sources. The photometric errors have been estimated based on the variability.}
  \label{fig:SMCSEDmodelfit}
\end{figure}

For the LMC and SMC we assume the star lies at a distance $d$ between $d_{\mathrm{min}}$ and $d_{\mathrm{max}}$ to account for variations in distance across each galaxy due to its inclined geometry \citep{vanDerMarel2001} which corresponds to a range of allowed luminosity scale factors for each population. For the globular clusters the allowed range in the scale factor for each source was determined, using the distances and associated error listed in Table~\ref{tab:GlobClusterSample}. Where possible the scale factor used for Galactic sources corresponds to its \emph{Hipparcos} distance if the relative errors are less than 20 per cent, otherwise we adopt a distance range of $d < 4$ kpc. Table \ref{tab:scale} lists the distance range and scaling factor allowed for each galaxy. As the luminosity is a fitting parameter to the distance, we attribute no significance to its best-fit values.  

To prevent any over-interpretation of the best fit parameters, we consider all sets of model parameters that produce a good fit to the source, with $\chi^2 - \chi_{\mathrm{best}}^{2} < 3$, where $ \chi_{\mathrm{best}}^{2}$ is the reduced \chisq ~value of the best-fitting model for each source, following the method employed by \cite{Robitaille2007}. For each parameter we find the average value and its standard deviation from the range of acceptable fits. Fig.~\ref{fig:SMCSEDmodelfit} shows an example best-fitting model to the SEDs of an O-AGB and a RSG in the SMC. The error bars are an indication of the expected change in flux due to variability. The IRS spectrum for each source is plotted to illustrate if a good fit has been achieved. These are not included in the fitting as the mineralogy is a fixed parameter in the {\sc grams} model grid.

The dust mass-loss rates determined through the spectral fitting of the SEDs with the {\sc grams} model grid are listed in Tables~\ref{tab:LMCsample}--\ref{tab:GlobClusterSample}. In principal, the fitting routine also provides uncertainties for the individual dust mass-loss rates, of order 10 to 55 per cent, however, in absolute terms, the errors are much greater than this due to the inherent uncertainties and assumptions in the model.

In order to assess the accuracy of the absolute values derived in the {\sc grams} fitting, we compare our derived dust mass-loss rates to those determined by \cite{Groenewegen2009} and \cite{McDonald2011} for common sources between the data sets. The {\sc grams} fits are systematically greater by a factor of 3--8. This is predominantly due to a difference in the choice of optical constants and grain shape. 
The {\sc grams} grid uses the astronomical silicates of \cite{Ossenkopf1992}, while both \cite{Groenewegen2009} and \cite{McDonald2011} use a combination of silicates, metallic iron and aluminium oxide. Different physical parameters used in modelling the dust envelope and central star also contribute to this discrepancy.  For more details on how the {\sc grams} models compare to other modeling efforts we refer the reader to \cite{Sargent2011}.

Our analysis regarding the onset of crystallinity is very sensitive to the uncertainties in the dust and gas mass-loss rates, thus it is important to account for the possible sources of error and caveats used in the modeling to prevent misinterpretation.
In the models we have assumed that all outflows have the same constant velocity of 10 kms$^{-1}$, whereas in reality they may have expansion velocities between $\sim$5 and $\sim$25 kms$^{-1}$, even within the same galaxy. However, by assuming a constant velocity the dust and gas mass-loss rates obtained in the fitting can be used as a proxy for the gas and dust column densities. As we have been consistent in our treatment of the samples and in the determination of the dust mass-loss rates, the relative uncertainty between the sources is small compared to the absolute uncertainties, allowing comparisons to be drawn across the sample. 
The derived gas mass-loss rates have an additional uncertainty because of the adopted scaling relation between dust and gas in circumstellar environments. For oxygen-rich evolved stars, this relation depend strongly on metallicity, however, it is not well constrained. As such, the absolute values for the gas mass-loss rate should be treated with a degree of caution, nevertheless, we expect the assumed gas-to-dust ratios to provide a reasonable first order approximation of the relative gas mass-loss rates.

\subsection{Measuring the crystalline silicate features} \label{sec:Featuremeasure}


\begin{table}
 \begin{minipage}{84mm}
 \caption{Wavelength intervals for extracting crystalline silicate emission features cf.~Fig.~\ref{fig:featureMeasure}.}
 \label{tab:featureExtraction} 
 \centering
 \begin{tabular}{@{}ccc}
   \hline
   \hline
   Feature ($\mu$m) &    Continuum interval ($\mu$m)  \\
\hline                                                                                           
    23.6          &    22.3 - 24.6                  \\
    28.1          &    26.0 - 31.1                 \\
    33.6          &    31.5 - 35.0                 \\
   \hline
 \end{tabular}
 \end{minipage}
\end{table}


\begin{figure}
\includegraphics[width=84mm]{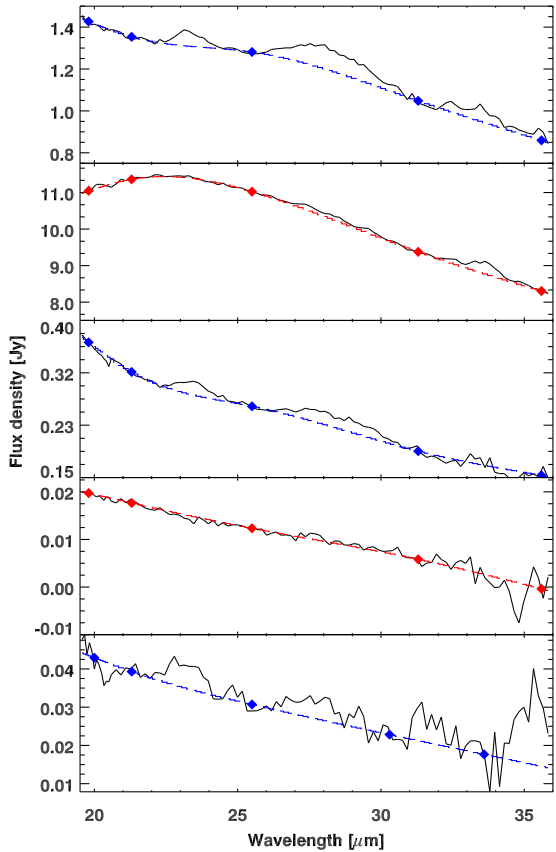}
 \caption{Example of the spline fitting for an LMC O-AGB (top), LMC RSG (2nd panel), SMC O-AGB (3rd panel), SMC RSG (4th panel). The spline (dashed line) is fitted to the spectra at the marked wavelengths. The top three spectra show crystalline emission features, while the spectrum in the 4th panel does not. The bottom panel shows SSTISAGEMC J053128.44$-$701027.1 (SSID 130), where alternative spline points were selected in-order to produce a smooth continuum. Fig.~\ref{fig:featureMeasure} presents the resulting continuum-divided spectra.}
  \label{fig:Splinefit}
\end{figure}

\begin{figure}
 \includegraphics[width=84mm]{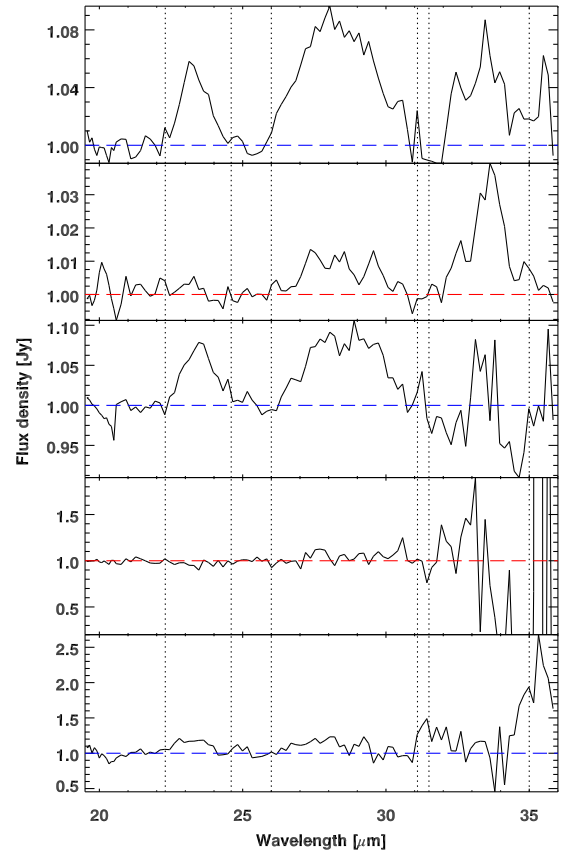}
 \caption{Continuum-divided spectra from the sources shown in Fig.~\ref{fig:Splinefit}. The dotted vertical lines mark the regions within which the complex strength was measured.}
  \label{fig:featureMeasure}
\end{figure}

Narrow spectral features due to crystalline silicates tend to cluster in well-defined complexes, found near 10, 18, 23, 28, 33, 40 and 60 $\mu $m \citep{Molster2002b}.  We do not extract information for the features near 10 and 18 $\mu$m as they are impossible to cleanly separate from the broader, underlying amorphous silicate features.

In order to determine the properties of crystalline silicate emission features across a range of metallicities, we have defined a continuum representing a featureless thermal dust component, which we used as a baseline to obtain an enhanced view of the spectral features beyond 20 $\mu$m. To produce a smooth continuum, a 5-pixel boxcar smoothing algorithm was applied to the IRS spectra to reduce the noise fluctuations between each data point (while preserving the resolution). However, it is possible that this may result in any low-contrast, broad dust features being treated as continuum. The continuum was determined by applying a cubic spline fit in $F_{\nu}$-space for $\lambda > 19.5 \, \mu$m. Spline points were fitted to the spectra at 19.8, 21.3, 25.5, 31.3, 35.6 \mum where the continuum is well defined and no strong spectral features are present. Examples of the spline fitting procedure can be seen in Fig.~\ref{fig:Splinefit}. In instances where a feature is present in a spectrum at a standard spline point (for example SSID130), the spline point was moved to the closest alternative wavelength where no feature could be visually identified. A similar procedure was applied to the {\em ISO} SWS spectra after regridding to the {\em Spitzer} IRS resolution. 

We prefer the above method compared to the alternate approach used by some authors, in which a local continuum is defined for each feature via linear interpolation between flux densities in a given wavelength range on either side of the feature. As slight variations in the position of the feature and noise in the spectra all have strong effects on the determination of the continuum because the fitting region is small. Also the dust continuum is inherently non-linear, with changes in curvature, producing a non-systematic offset between sources.


We determine equivalent widths of the 23-, 28- and 33-$\mu$m features in each spectrum to provide a quantitative measure of their strengths. These are calculated with respect to the underlying continuum as determined by the aforementioned spline fitting process. The equivalent widths ($W_{\rm eq}$) can then be determined by:
\begin{equation}
W_{\rm eq} = \sum_{\lambda} \left( 1 - \frac{F_{\rm obs}}{F_{\rm cont}} \right) {\rm d}\lambda ,
\end{equation}
where $F_{\rm obs}$ and $F_{\rm cont}$ are the observed flux and the flux of the spline-fitted continuum. The wavelengths used to define the blue and red sides of each feature are given in Table \ref{tab:featureExtraction}. An example of the feature extraction process is given in Fig.~\ref{fig:featureMeasure}. 

If a feature is particularly weak compared to the continuum, its strength may be influenced by how the continuum is fit. To quantify this effect, we also determined equivalent widths using linear interpolation to approximate the continuum under the features. For spectra with high contrast features or with a good S/N ratio the difference in strength is of the order of $\lesssim$ 15 per cent, however, there is a large disparity (up to $\sim$65 per cent) in strengths for low contrast features.

Both the spline and linear interpolation methods will underestimate the total contribution from the crystalline silicate components to the total flux. These methods fit the continuum under each crystalline silicate emission feature, however, the crystalline silicates grains will also contribute some flux to the underlying continuum which is composed of contributions from all dust species of varying mineralogy, lattice structure, temperature and size and shape distributions.

\section{Results} \label{sec:results}

\subsection{Mass-loss rates}\label{sec:MLRs}

\begin{table*}
 \caption{Summary of mean mass-loss rates.}
 \label{tab:statsMLR} 
 \centering
 \begin{tabular}{@{}l@{\ }lccccccc@{}}
   \hline
   \hline
     &        &     \multicolumn{3}{|c|}{Sources with only}        &  \multicolumn{3}{|c|}{Sources with}                \\   
     &        &     \multicolumn{3}{|c|}{amorphous silicates}      &  \multicolumn{3}{|c|}{crystalline silicates}        \\    
\multicolumn{2}{|c|}{Population} &$ \langle \log {\dot{M}_{dust}} \rangle$ &   $\langle \log \dot{M}\rangle$ & N &   $ \langle \log {\dot{M}_{dust}}  \rangle$ & $ \langle \log \dot{M} \rangle$ & N  \\	
   \hline 
MW             &  O-AGB  &  $-$8.07     &  $-$6.07   &   59     &   $-$6.87     &  $-$4.87      &      50       \\ 
MW             &  RSG    &  $-$7.95     &  $-$5.95   &   15     &   $-$7.41     &  $-$5.41      &       7       \\ 
LMC            &  O-AGB  &  $-$8.19     &  $-$5.89   &   36     &   $-$7.20     &  $-$4.90      &      18       \\ 
LMC            &  RSG    &  $-$8.21     &  $-$5.91   &   49     &   $-$7.53     &  $-$5.23      &      11       \\ 
SMC            &  O-AGB  &  $-$9.18     &  $-$6.49   &   10     &   $-$7.42     &  $-$4.72      &       5       \\ 
SMC            &  RSG    &  $-$8.83     &  $-$6.13   &   15     &   $-$7.49     &  $-$4.78      &       1       \\ 
Terzan 5       &  O-AGB  &  $-$8.06     &  $-$5.98   &    4	&   $-$7.64	&  $-$5.56      &       1	\\	
Palomar 10     &  O-AGB  &  $-$8.05	&  $-$5.95   & 	  1	&   \dots	&  \dots	&	0	\\	
NGC 5927       &  O-AGB  &  $-$7.96	&  $-$5.61   & 	  1	&   $-$8.21	&  $-$5.86	&	1	\\	
NGC 6760       &  O-AGB  &  $-$8.19	&  $-$5.71   & 	  2	&   \dots	&  \dots	&	0	\\		
NGC 6356       &  O-AGB  &  $-$8.66	&  $-$6.16   &    4	&   \dots	&  \dots	&	0	\\		
NGC 6441       &  O-AGB  &  $-$8.87	&  $-$6.31   &    2	&   \dots	&  \dots	&	0	\\	
NGC 6388       &  O-AGB  &  $-$8.83	&  $-$6.26   &    2	&   \dots	&  \dots	&	0	\\	
NGC 6637       &  O-AGB  &  $-$8.32	&  $-$5.66   &    2	&   \dots	&  \dots	&	0	\\		
NGC 6352       &  O-AGB  &  \dots	&  \dots     &    0	&   $-$8.49	&  $-$5.80	&	1	\\		
IC 1276	       &  O-AGB  &  $-$8.36	&  $-$5.67   &    2	&   \dots	&  \dots	&	0	\\	
NGC 6838       &  O-AGB  &  $-$8.63	&  $-$5.90   &    1	&   \dots	&  \dots	&	0	\\		
NGC 6712       &  O-AGB  &  $-$8.64	&  $-$5.67   &    2	&   \dots	&  \dots	&	0	\\		
NGC 362	       &  O-AGB  &  $-$8.99	&  $-$5.79   &    2	&   \dots	&  \dots	&	0	\\	
NGC 5139       &  O-AGB  &  $-$9.66	&  $-$6.03   &   10	&   $-$8.36	&  $-$4.73	&	1	\\
   \hline
 \end{tabular}
\end{table*}

\begin{figure*}
  \centering
  \subfloat[${\dot{M}_{dust}}$ distribution]{\label{fig:DMLR_boxnwisk}\includegraphics[width=84mm]{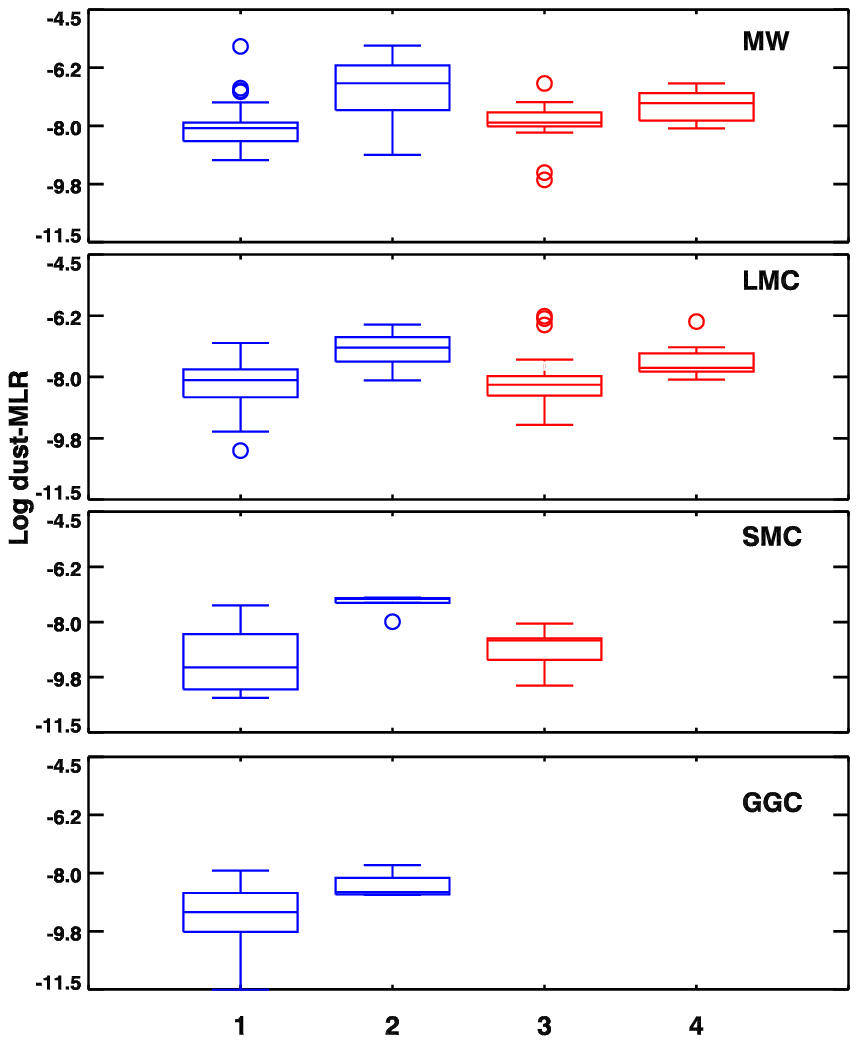}}
  \subfloat[$\dot{M}$ distribution]{\label{fig:TMLR_boxnwisk}\includegraphics[width=84mm]{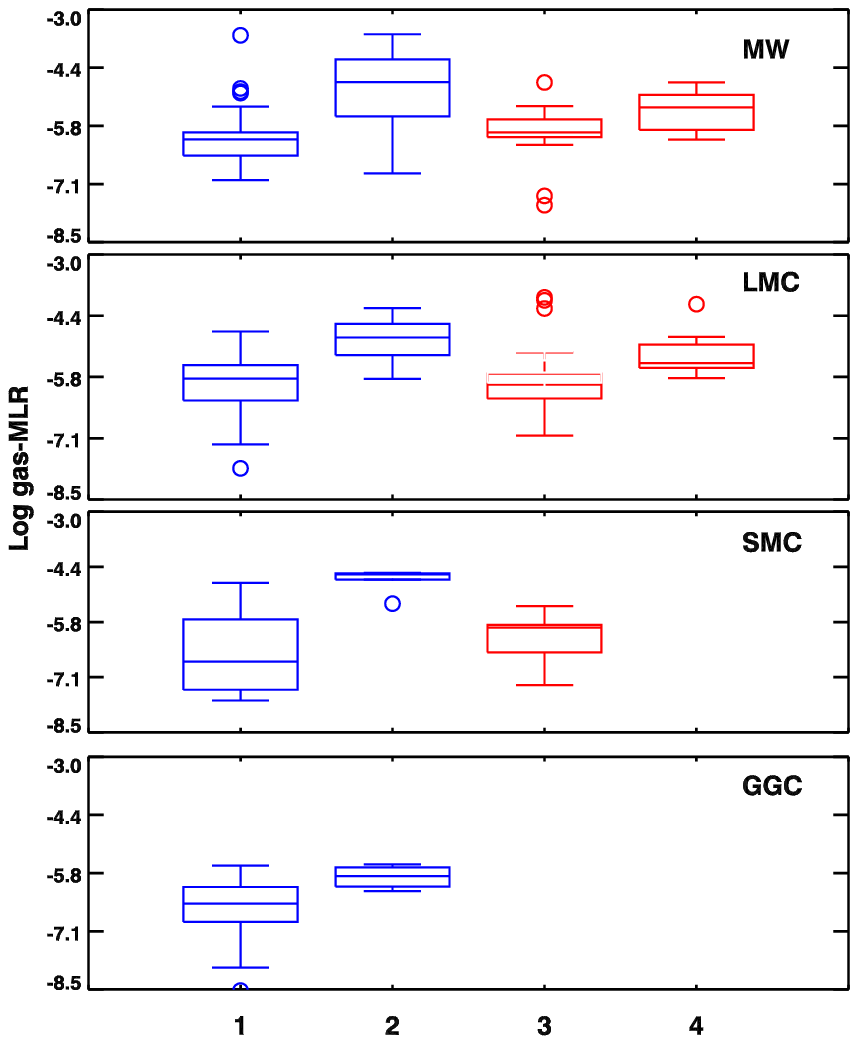}}
  \caption{Overview of the dust and gas mass-loss rate distributions for each population. Column 1: O-AGBs with solely amorphous silicates, column 2: O-AGBs with crystalline silicates, column 3: RSGs with only amorphous silicates, column 4: RSGs with crystalline silicates. The box encloses the inter-quartile range, the whiskers extend out to 1.5 times the lower or upper quartile or to the maximum or minimum value of the data (if smaller).  Outliers are indicated by circles.}
  \label{Fig:MLRstats}
\end{figure*}

We have separated the O-AGB and RSGs stars in each galaxy in order to examine the influence of the mass-loss process on the dust production in these two classes of object separately. Furthermore, we distinguish between sources in which crystalline silicates are visually apparent and those whose silicates appear to be solely amorphous. For each population we determine the mean $\log {\dot{M}_{dust}} $ and $\log \dot{M} $. In general, the dust mass-loss rate for the sources with only amorphous silicates increases as the metallicity of the sample increases. Thus, the SMC and globular cluster sources have low dust mass-loss rates while the Galactic field sources have the largest dust mass-loss rates. The individual objects within a galaxy encompass a range of mass-loss rates as indicated in Fig.~\ref{Fig:MLRstats}.  An overview of $ \langle \log {\dot{M}_{dust}} \rangle$ and $\langle \log \dot{M}\rangle$ for each population is presented in Table~\ref{tab:statsMLR}. 

In each instance the mean dust and gas mass-loss rate is higher for the sources which exhibit crystalline silicate features compared to those that do not. Also, while a low mass-loss rate is no guarantee that the silicates will be entirely amorphous, it is evident that high mass-loss rate sources have a greater disposition to exhibit crystalline silicate features. The mean dust mass-loss rates for the sources with crystalline silicates do not show the same metallicity dependence as the  sources with only amorphous silicates. As these objects have different gas densities but similar dust column densities this may indicate that crystallinity depends on the dust density in the wind. 
The mass of the central star may also be significant: a greater fraction of O-AGB stars (56 per cent) display crystalline bands compared to the RSGs (24 per cent) and, as the metallicity decreases, the proportion of RSGs with crystalline features also drops. This may be due to differences in the silicate dust composition of RSG and O-AGB stars, with the former exhibiting amorphous Ca-Al-rich silicates before forming Mg-rich silicates \citep{Speck2000, Verhoelst2009}.

\begin{figure}
  \includegraphics[width=84mm]{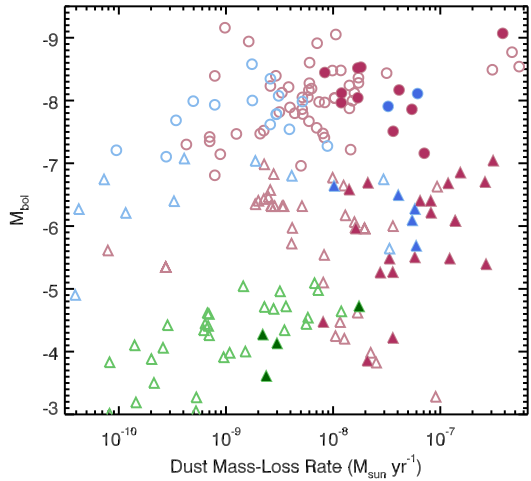}
  \caption{Bolometric luminosity versus dust mass-loss rate for the LMC (red colours), SMC (blue) and globular cluster (green) sources. O-AGB stars are plotted as triangles and RSG as circles. Objects with crystalline silicates are plotted with closed symbols. Milky Way sources are not included because of distance uncertainties.}
  \label{Fig:MLR_Mbol}
\end{figure}

Fig.~\ref{Fig:MLR_Mbol} shows the bolometric magnitude as a function of dust mass-loss-rate. We detect crystalline silicates across the full AGB and RSG luminosity range, from the early AGB to the most luminous RSGs. This indicates that we are sensitive to crystalline features down to a low luminosity limit, not just in the brightest sources where there is better contrast between the feature and the noise. For the Magellanic Clouds our sample is biased towards the brightest AGBs and RSGs or dusty sources with high mass-loss rates. Here the threshold density for the onset of crystallinity is greater than that of the Galactic field stars and globular cluster sources. As such, the detection of crystalline features in the Magellanic Clouds may be limited by the signal-to-noise.

\subsection{Mass-loss rates and crystalline feature strength}\label{sec:MLRxSIlresults}
 
\begin{figure*}
  \includegraphics[trim = 0mm 12.5mm 0mm 0mm, clip]{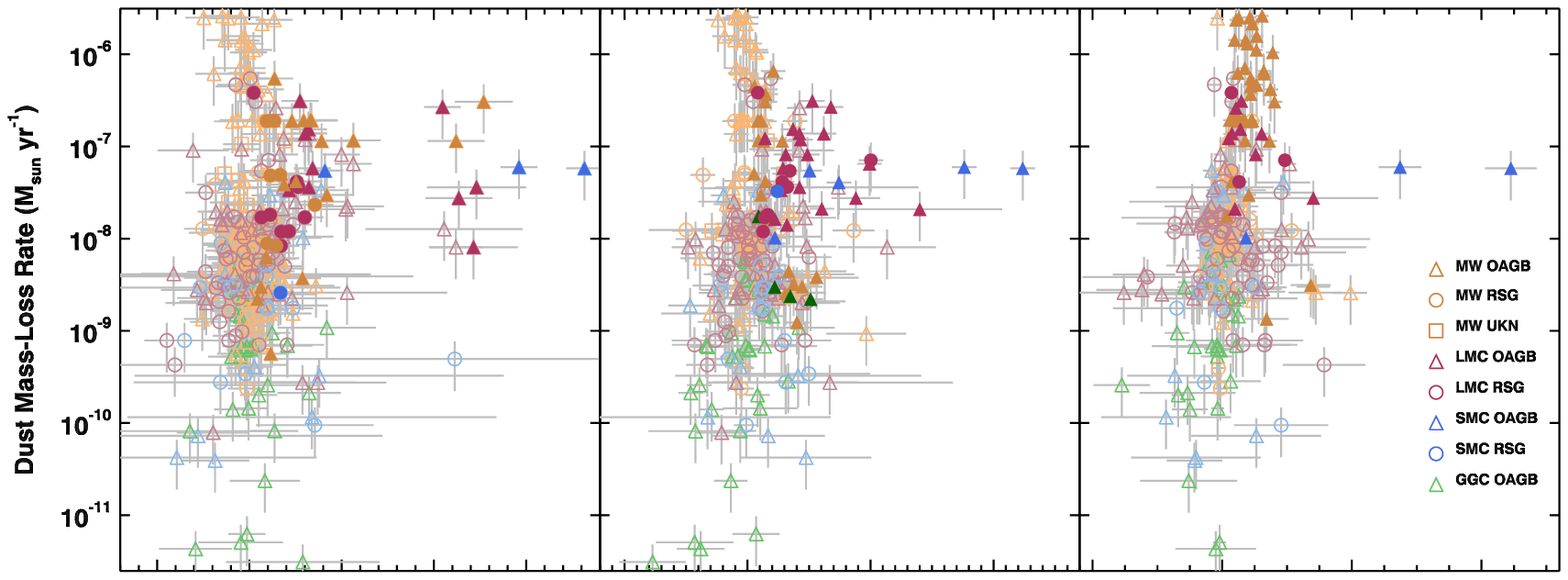}
  \includegraphics[trim = 0mm 0mm 0mm 5mm, clip]{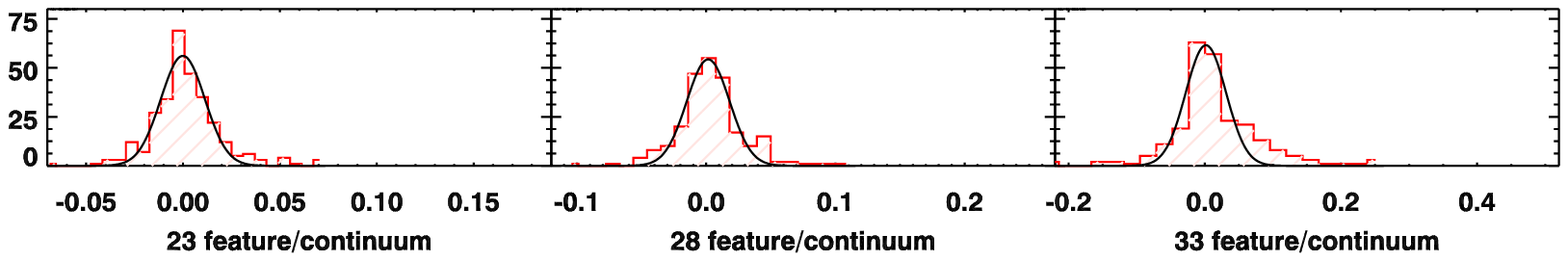}
  \includegraphics{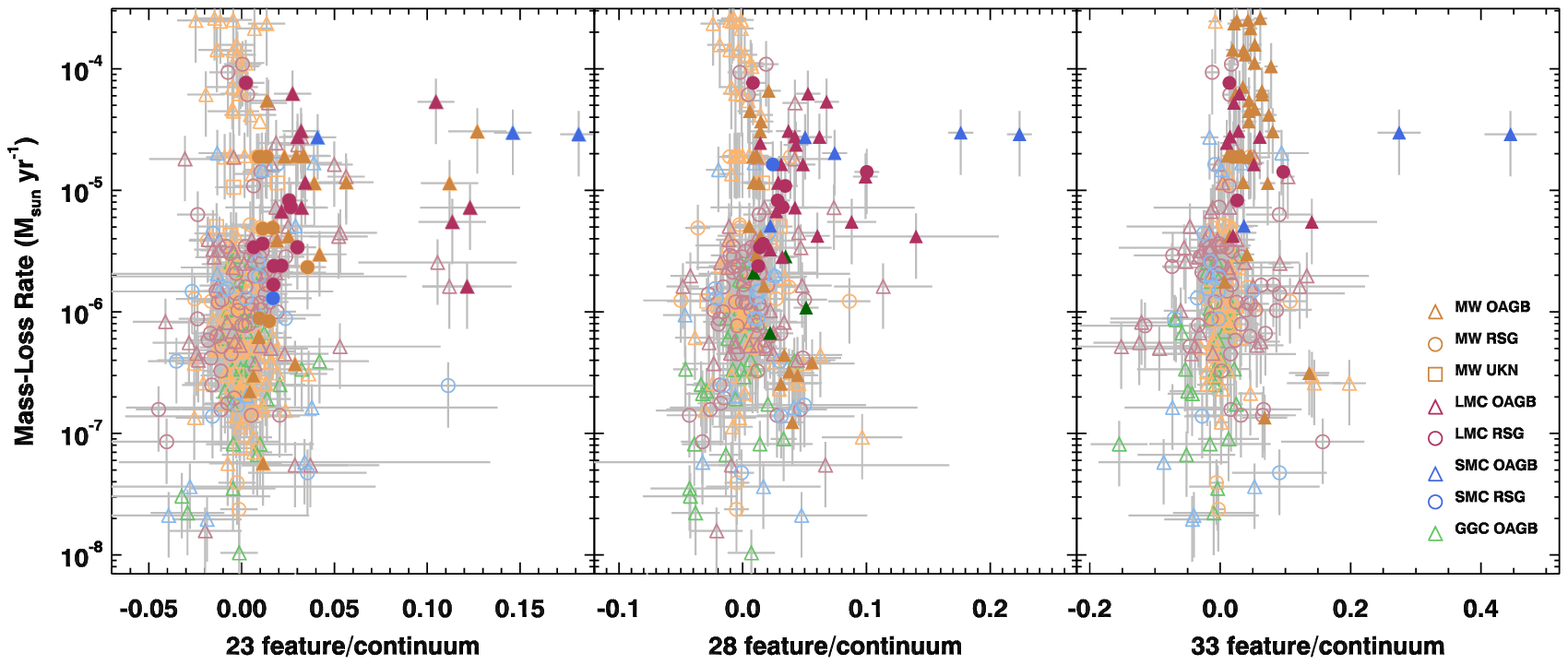}
   \caption{\emph{Top:} The strength of the 23-, 28- and 33-\mum features versus ${\dot{M}_{dust}}$, the dust mass-loss rate. O-AGB stars are represented by triangles, RSG as circles and sources which cannot be distinguished as either O-AGB or RSG are indicated by squares. The LMC sources are plotted in red; SMC: blue; MW: orange; and the globular clusters: purple. Filled symbols indicate that a crystalline feature can be visually identified. For sources where no feature is detected, a Gaussian is fitted to the distribution of feature strength/continuum.
 \emph{Bottom:} The strength of the 23-, 28- and 33-\mum features versus $\dot{M}$, the total mass-loss rate. As with the dust mass-loss rates there is no apparent threshold value between sources with and without 23- and 28-$\mu$m silicate features.}
  \label{Fig:MLRvft}
\end{figure*}

\begin{figure}  \includegraphics[width=84mm]{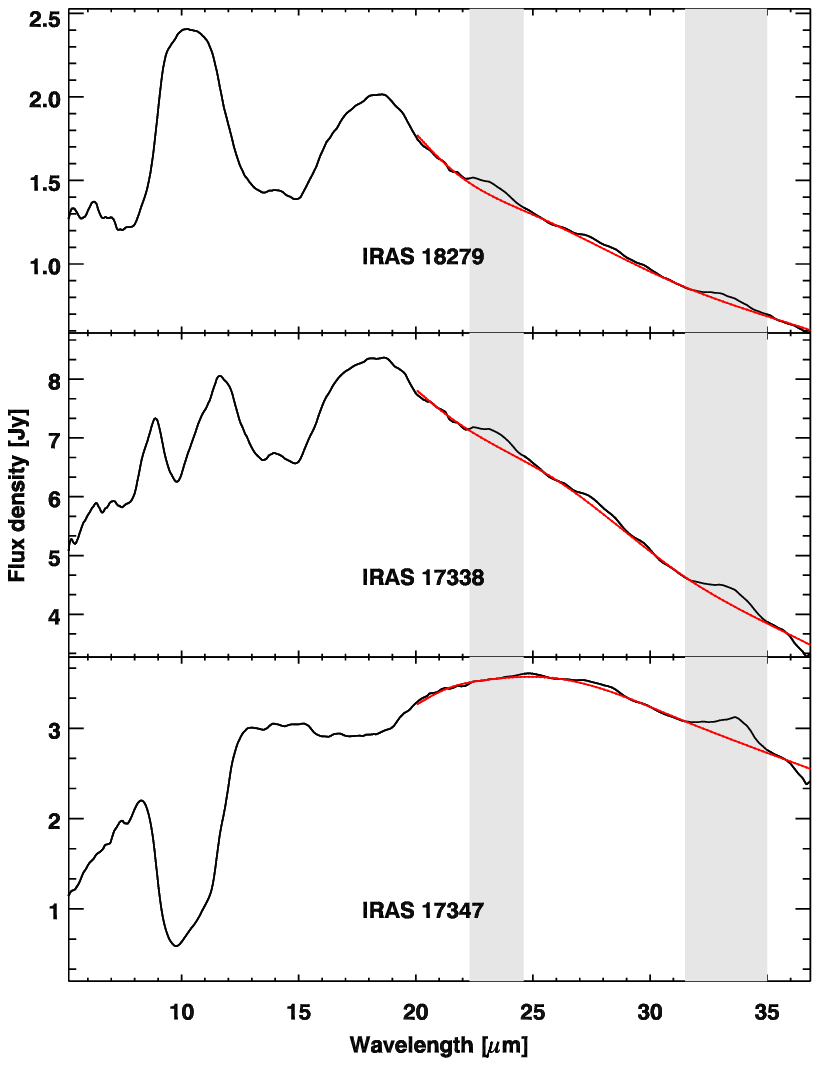}
  \caption{Three {\em Spitzer} spectra ordered in terms of the 10 \mum optical depth. As the dust shell becomes optically thick the 23-\mum feature strength becomes suppressed. The smooth red lines show the adopted continua (spline). The shaded regions indicate the wavelengths where the 23- and 33-\mum features were measured.}
  \label{Fig:Spec23supress}
\end{figure}

Fig.~\ref{Fig:MLRvft} shows the strengths of the 23-, 28- and 33-\mum features against the dust mass-loss rate (upper panel) and total mass-loss rate (lower panel). In both instances, objects with crystalline silicate features are represented by filled symbols, while open ones represent sources where crystalline silicates are not visually apparent. It is interesting to compare and contrast the behaviour of the crystalline features at the different wavelengths, in particular the 23- and 33-\mum features, which have the same principal carrier, forsterite. At 33 $\mu$m, features due to the the crystalline silicate grains only become apparent at the highest mass-loss rates, where the majority of the sources display some crystallinity. This is usually interpreted as evidence that there is a threshold value in the mass-loss rate at which crystalline silicates can form \citep{Cami1998, Sylvester1999}. We observe, however, that the forsterite feature at 23 \mum paints a different picture. Here, the transition between sources which exhibit a feature and those which do not is blurred, indeed at the highest mass-loss rates ($\dot{M}$ $>$ $10^{-4}\, {\rm M_{\odot}\, yr^{-1}}$) the detection rate of the 23-\mum feature drops: this may be a consequence of the forsterite grain temperature or it may indicate that the 23-\mum band is going into absorption due to high column density. An example of the changing relative strengths of the two forsterite features as the optical depth of the shell increases is plotted for three MW stars in Fig.~\ref{Fig:Spec23supress}. Like the 23-\mum forsterite feature, the 28-\mum enstatite feature spans a wider mass-loss rate range than the 33-\mum feature. However, in all three cases no evidence for crystallinity is seen in the lowest mass-loss rate sources. 

The sources where no feature is visually apparent are well described by a Gaussian distribution around a strength/continuum ratio of zero. This implies that this spread is due to stochastic noise in the spectra and that we can trust systematics in the crystalline silicate detections. For the 23- and 33-\mum bands there is a slight positive skew, this is probably due to real features which we are unable to visually identify in the data as a consequence of substantial noise in the \emph{Spitzer} spectra at these wavelengths. 
The 33-\mum band can be susceptible to contamination from diffuse [S~III] and [Si~II] emission lines, which may be improperly subtracted due to their non-uniform variation across the slit. We have taken care to minimise this contamination by excluding this region from our analysis if we detect the presence of sky lines. Consequently, the detection rate of the complexes represents a lower limit. 

\subsection{Dust temperature}

\begin{figure}
  \includegraphics[width=84mm]{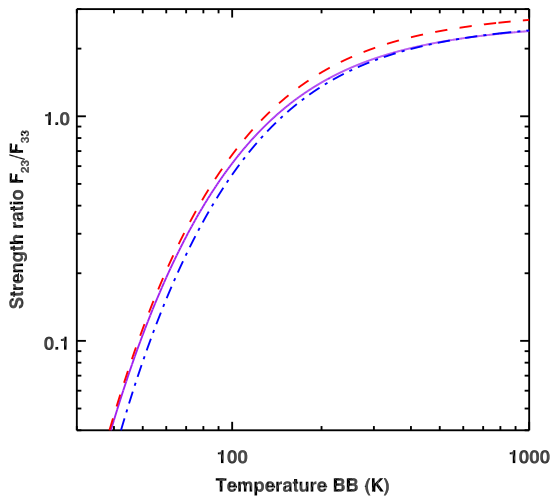}
   \caption{The ratio of relative peak strengths above continuum for the 23- and 33-$\mu $m forsterite bands as a function of the forsterite dust grain temperature. The different curves represent the different forsterite laboratory data: solid purple: \protect\citep{Koike1999}; dashed red: \protect\citep{Jager1998} continuous distribution of ellipsoids shape distribution (CDE); and  dash dot blue: \protect\citep{Jager1998} homogeneous spherical particle shape distribution (Mie).}
 \label{Fig:Temp_diagnostic}
\end{figure}

The strengths of individual bands are affected by the abundance of each species, and by grain shape and temperature. By eliminating the former two parameters, we can derive the temperature of the emitting grains. We can eliminate the need to parametrise the abundance by taking the strength ratio of features which have the same carrier: in the case of forsterite, we can use the ratio of the 23- and 33-$\mu$m features.

 Eliminating the effects of grain shape are more difficult, as these are further confounded by the choice of laboratory optical constants used to create the feature shapes. Emission models for the forsterite grains were computed from laboratory data \citep{Jager1998,Koike1999}. Both these datasets give a good match to the observed feature profiles and do not differentiate different crystallographic axes, reducing the number of parameters which we must fit. From these, we use three sets of absorption efficiencies ($Q_{\rm fors}(\nu)$). The first comes directly from \citet{Koike1999}. The second is derived from the optical constants of \citet{Jager1998}, using homogeneous spherical particles with radii of 0.01 $\mu$m via Mie theory. The third is similarly calculated using a continuous distribution of ellipsoid grains with equivalent volumes. 

Assuming that the long wavelength ($\lambda > 21~\mu$m) forsterite features are formed in an optically-thin region of the outflow then the crystalline forsterite spectrum can be approximated by: 
\begin{equation}  \label{equ:Fostcomp} F(\nu)_{{\rm fors}} \sim B(T_{{\rm fors}},\nu) \times  Q(\nu)_{{\rm fors}} + F_{{\rm cont}} \end{equation}
where $B(T_{{\rm fors}},\nu) $ is a blackbody of temperature $T_{{\rm fors}}$,  $Q(\nu)_{{\rm fors}}$ is the forsterite absorption efficiency, and  $F_{{\rm cont}}$ represents the sum of the featureless thermal dust component and stellar continuum. This component is effectively removed from the spectra by subtracting the spline-fitted continuum we use in Section~\ref{sec:Featuremeasure}.  We therefore also subtract an identically-fitted spline from the modelled forsterite spectrum to ensure consistency.

These continuum-subtracted forsterite absorption efficiencies were then multiplied by blackbody functions corresponding to temperatures between 30 and 1000~K at 5~K intervals. The strengths of each feature were then determined by summing the remaining flux over the wavelength intervals given in Section~\ref{sec:Featuremeasure}, and a ratio of the 23- to 33-$\mu$m features was determined. Fig.~\ref{Fig:Temp_diagnostic} shows how this ratio changes with temperature, allowing direct comparison to observed band strengths. An average of the resulting forsterite temperatures are given in Table~\ref{tab:forsteriteTemp}. In sources where there is no 23-\mum feature detection, T$_{fors}$ may be incorrect due to self-absorption. This approach relies on the ratio of features' strengths, it is not greatly affected by the change in feature position due to temperature, and can therefore be treated as a robust estimate of temperature once one takes into account the above caveats regarding grain shape and self-absorption.

\begin{figure}
\includegraphics[width=84mm]{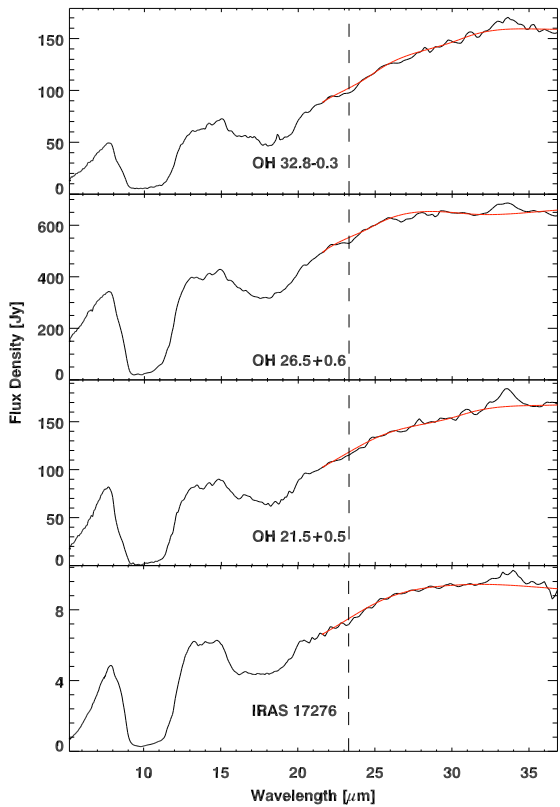}
\caption{Spectra showing crystalline silicate absorption and emission features. To enhance the visibility of the crystalline features in the spectra the adopted continua are shown (smooth red lines). The position of the 23-\mum feature is indicated by a dashed line.}
\label{Fig:xsilAbs}
\end{figure}

In the 23-\mum wavelength regime, the suppression of the forsterite feature for the majority of the sources where crystallinity is present is probably a consequence of the poor contrast of the cool grains above the amorphous dust continuum and, as such, they are indiscernible with \emph{Spitzer} and \emph{ISO}. For the sources with extremely high mass-loss rates we believe the absence of the 23-\mum forsterite feature is due to self-absorption. Fig.~\ref{Fig:xsilAbs} shows the spectra of four Galactic O-AGB stars undergoing intense mass-loss rates. In these sources, amorphous silicate absorption features are seen at 10 and $\rm {20{\ }\mu}$m; OH 32.8$-$0.3,  OH 26.5$+$0.6 and IRAS~17276$-$2846 also clearly exhibit crystalline silicate absorption features up to $\rm {33{\ }\mu}$m where the feature appears in emission. The extreme OH/IR star IRAS~16342$-$3814 is also known to have some crystalline silicate features in absorption \citep{Dijkstra2003}. This object is undergoing an intense mass-loss of $\sim10^{-3}\, {\rm M_{\odot}\, yr^{-1}}$, which is approximately an order of magnitude greater than the highest mass-loss rates determined in this work.
This behaviour is similar to that modeled by \cite{deVries2010}, who found that at mass-loss rates greater than $10^{-5}~$M$_{\odot}~{\rm yr}^{-1}$ even the 33-$\mu $m band becomes self-absorbed in some lines of sight. These results show that there is a limit to the strength of the crystalline silicates in AGB stars.

\begin{table} 
 \begin{minipage}{84mm}
 \caption{Derived forsterite temperatures for stars with a 33-$\mu$m feature.  In sources where there is no 23-\mum feature detection, T$_{{\rm fors}}$ may be incorrect due to self-absorption.} 
 \label{tab:forsteriteTemp} 
 \centering
 \begin{tabular}{@{}lc@{ \ \ }c@{}}
   \hline
   \hline
    Star           & 23 $\mu$m Ft.   & T$_{{\rm fors}}$ (K) \\
   \hline                                               
NML Cyg             & Y   &    88   $\pm$  10    \\ 
NML Cyg             & Y   &   121   $\pm$  23    \\
WX Psc              & Y   &    89   $\pm$  10    \\ 
WX Psc              & Y   &    89   $\pm$  10    \\ 
IRC $+$50137        & Y   &   170   $\pm$  47    \\ 
HV 12956            & Y   &   102   $\pm$  13    \\ 
OH 104.91$+$2.41    & Y   &    67   $\pm$  6     \\ 
IRAS 03434$+$5818   & Y   &   458   $\pm$  365   \\
IRAS 04553$-$6825   & Y   &    73   $\pm$  7     \\ 
IRAS 04545$-$7000   & Y   &    88   $\pm$  10    \\ 
IRAS 05402$-$695    & Y   &    98   $\pm$  12    \\ 
IRAS 05298$-$6957   & Y   &   108   $\pm$  15    \\ 
IRAS 05389$-$6922   & Y   &   498   $\pm$  390   \\
IRAS 135811$-$5444  & Y   &   193   $\pm$  63    \\ 
IRAS 17304$-$1933   & Y   &   321   $\pm$  190   \\ 
IRAS 17338$-$2140   & Y   &   169   $\pm$  46    \\ 
IRAS 17413$-$3531   & Y   &   158   $\pm$  39    \\ 
IRAS 18279$-$2707   & Y   &   507   $\pm$  385   \\ 
IRAS 18291$-$2900   & Y   &   245   $\pm$  108   \\ 
IRAS 19256$+$0254   & Y   &   546   $\pm$  364   \\ 
IRAS 19456$+$1927   & Y   &   267   $\pm$  130   \\ 
MSX SMC 134         & Y   &   117   $\pm$  19    \\
MSX LMC 807         & N   &  $ \le  $  150 \\ 
S Scl               & N   &  $ \le  $   64 \\ 
HD 269599           & N   &  $ \le  $   66 \\ 
AFGL 2199           & N   &  $ \le  $   70 \\ 
AFGL 2403           & N   &  $ \le  $   53 \\ 
AFGL 2403           & N   &  $ \le  $   68 \\ 
AFGL 2374           & N   &  $ \le  $   89 \\ 
AFGL 5379           & N   &  $ \le  $   47 \\ 
AFGL 5379           & N   &  $ \le  $   46\\ 
AFGL 5535           & N   &  $ \le  $   50 \\ 
OH 21.5$+$0.5       & N   &  $ \le  $   52 \\ 
OH 26.2$-$0.6       & N   &  $ \le  $   43 \\ 
OH 26.5$+$0.6       & N   &  $ \le  $   63 \\
OH 32.8$-$0.3       & N   &  $ \le  $   45 \\
OH 127.8 $+$0.0     & N   &  $ \le  $   84 \\
OH 127.8 $+$0.0     & N   &  $ \le  $   56 \\
IRAS 05128$-$6455   & N   &  $ \le  $   70 \\ 
IRAS 04407$-$7000   & N   &  $ \le  $  122 \\
IRAS 08425$-$5116   & N   &  $ \le  $   56 \\ 
IRAS 05329$-$6708   & N   &  $ \le  $  206 \\ 
IRAS 05558$-$7000   & N   &  $ \le  $  153 \\ 
IRAS 17004$-$4119   & N   &  $ \le  $   56 \\ 
IRAS 17010$-$3840   & N   &  $ \le  $   51 \\
IRAS 17030$-$3053   & N   &  $ \le  $   53 \\ 
IRAS 17276$-$2846   & N   &  $ \le  $   54 \\ 
IRAS 17347$-$2319   & N   &  $ \le  $   47 \\ 
IRAS 17513$-$3554   & N   &  $ \le  $   84 \\ 
IRAS 18195$-$2804   & N   &  $ \le  $   60 \\ 
IRAS 18231$+$0855   & N   &  $ \le  $  513 \\ 
MSX SMC 024         & N   &  $ \le  $   84 \\ 
   \hline
 \end{tabular}
 \end{minipage}
\end{table}

\subsection{Feature shapes \& positions} \label{sec:avFt}

\begin{figure*}
\includegraphics{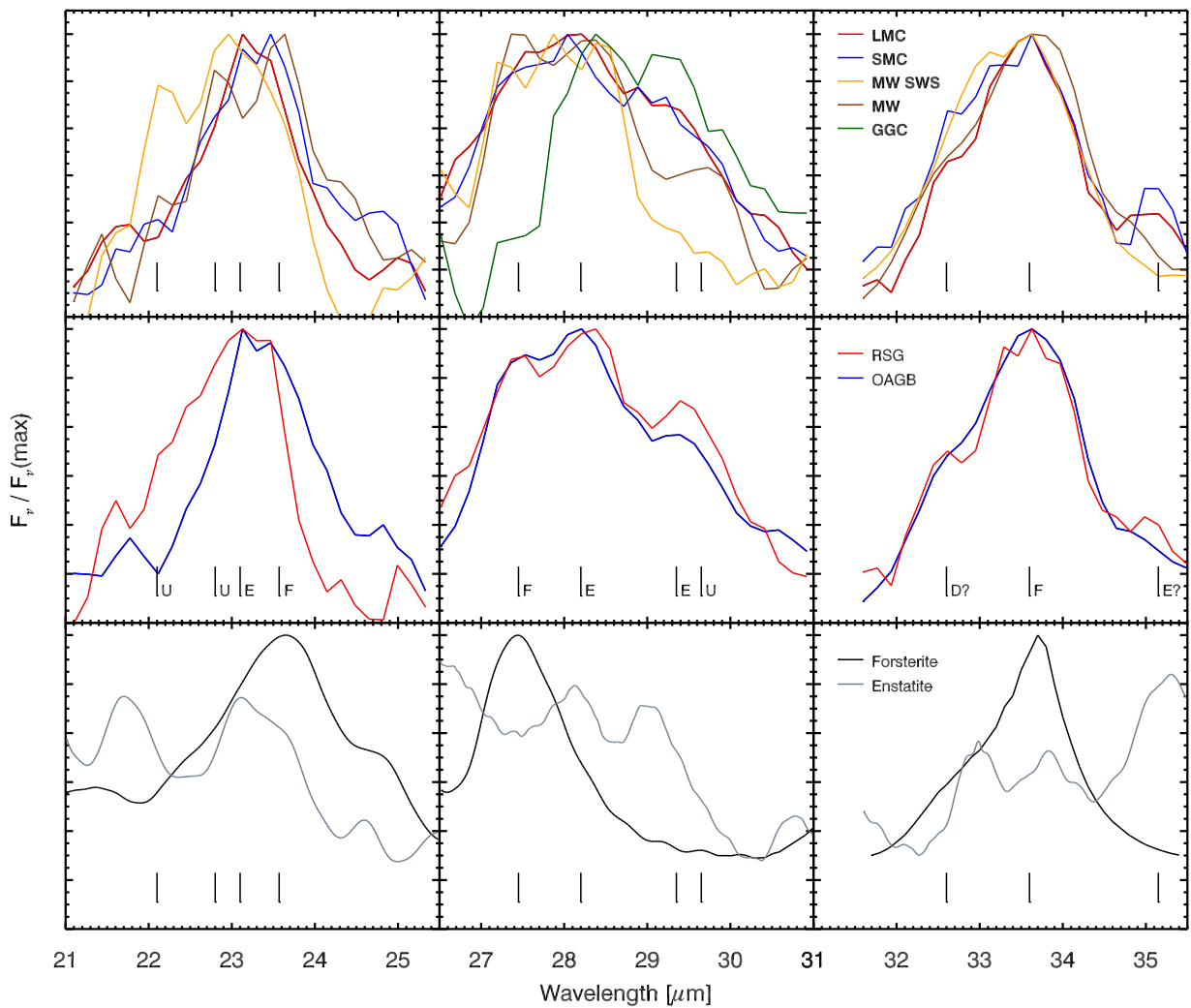}
\caption{Overview of the normalised mean continuum-subtracted 23-, 28- and 33-\mum  complexes. {\em Top}: the LMC, SMC, globular cluster and MW \emph{Spitzer} complexes are plotted in blue, red, green and brown respectively, while the re-gridded MW spectra from the SWS are plotted in orange. {\em Middle row}: the combined mean complex spectrum of O-AGB (blue) and RSG (Red) sources. The tick marks indicate the wavelengths where crystalline features are found. The crystalline species are indicated by a F (forsterite), E (enstatite), D (diopside), while a U indicates an unidentified species. {\em Bottom}: Laboratory spectra of forsterite (black) and enstatite (grey) are plotted for comparison \protect\citep{Koike1999}. }
\label{Fig:avFtshape}
\end{figure*}

To compare the crystalline features across the Milky Way, Magellanic Clouds and the Galactic globular clusters we derive a mean continuum-subtracted spectra for the complexes at 23, 28 and 33 $\mu$m. The emission from the crystalline silicate features can be isolated from the observed spectrum by subtracting a continuum according to Equation~\ref{equ:Fostcomp}. Emission features were extracted from the resulting continuum-subtracted spectra over the wavelength ranges specified in Table~\ref{tab:featureExtraction}, and averaged using a weighting factor (calculated from the point-to-point RMS) to account for the varying signal-to-noise in each region. The mean spectral complexes for the MW, LMC and SMC sources are shown in Fig.~\ref{Fig:avFtshape}. In the globular cluster sample only the  28-\mum spectral complex is present.

The precise shape and position of the continuum-subtracted features will depend on the spline-fitted continuum. In order to evaluate how sensitive a feature is to the defined continuum, we also computed average profiles using a linear continuum. The central wavelength of the 23-\mum feature (measured by finding the wavelength which bisects the integrated flux of the feature) has a mean variation of 0.08 \mum when a linear continuum was fitted, a mean change in position of 0.19 \mum was recorded for the 28-\mum feature and the 33-\mum feature had a 0.17 \mum change. In extreme cases where the features are weak compared to the continuum, or the spectra have a low signal-to-noise the central wavelength may move by 0.43 $\mu$m. For most cases the change in wavelength is relatively minor.  The average feature shape is remarkably insensitive to the method of continuum fitting, due in large part to the weighting factor and normalisation.

\subsubsection{The 23-\mum complex}

The 23-\mum complex is clearly seen in several of the stars across a wide range of dust mass-loss rates. However, it becomes less prominent and even absent in the highly-enshrouded sources. This band is entirely absent in the globular cluster sources. The complex is thought to be dominated by forsterite emission \citep{Molster2002b, Sloan2008}, with its peak fitting the dust spectra better than Fe-rich members of the olivine series. The average central wavelength of the complex, for each population, is around 23.35 $\mu $m, however, the variations between samples and individual sources is larger than that of the 33-$\mu $m feature.  

The complex is comprised of several blended features at 22.3,  22.8, 23.0 and 23.7 $\mu $m whose relative strengths provide the main difference between the samples. While the 22.3, 23.0 and 23.7 \mum features have previously been reported by \cite{Molster2002b}, this is the first-time the 22.8 $\mu $m component is seen. The complex is dominated by the 23.7-$\mu $m forsterite band, with the 23.0-$\mu $m enstatite feature also contributing to the substructure. The 22.3-$\mu$m and the newly discovered 22.8 $\mu$m bands have yet to be attributed to a dust species. 

The strength of the 23.0-\mum feature compared to the 23.7-\mum feature provides the main difference between the Magellanic Cloud mean profiles and the MW profiles. 
The mean profile of the SWS Galactic sources exhibits the largest deviation from the other mean spectra: this mean spectrum shows prominent 22.3- and 22.8-\mum features which are more subdued in the other profiles. Furthermore, the mean SWS Galactic spectrum is shifted towards the blue. 
The mean profile of the RSGs has a similar shift to shorter wavelengths with a central wavelength of $\sim$23.0 $\mu $m (compared to the O-AGB sources at 23.4 $\mu $m), as seen in the middle left panel of Fig.~\ref{Fig:avFtshape}.

\subsubsection{The 28-\mum complex} 

The 28-$\mu $m complex shows clear variations in the spectra with metallicity. Although the samples show a similar peak position around 27.8 $\mu $m, the spectral shape is different for the Magellanic Cloud, globular cluster and Galactic sources. This complex is comprised of several blended features of which the 27.5-$\mu $m forsterite feature, combined with the  28.2- and  29.4-$\mu $m enstatite features, are the main components \citep{Molster2002b}. In the Magellanic Cloud profiles, the broader feature may indicate an enhanced enstatite contribution compared to forsterite. There may also be a significant contribution from the unidentified 29.6 $\mu $m feature, which is especially pronounced in the MW {\em Spitzer} mean complex. Unlike the MW and the Magellanic Cloud profiles, the mean spectrum for the globular clusters does not show a forsterite component; here, the complex is dominated by enstatite.  There is no discernible difference between the O-AGB stars and the RSGs.

Determining an average profile for the MW SWS spectra was hampered by the troublesome boundary between bands 3D, 3E and 4 in this region, where known light leaks and spectral response calibration problems result in large uncertainties in the flux \citep{Sloan2003}. Several SWS spectra show a broad plateau extending from $\sim 27.5{-}32.2~\mu$m. We could find no evidence for a similar feature shape in any of the {\em Spitzer} spectra and we conclude that this plateau is probably an instrumental artifact. We have taken care to exclude any sources where this artifact may be present in the {\em ISO} SWS data before determining a mean spectrum for this sample. 

There is a large discrepancy between these mean profiles and the 28 $\mu $m outflow complex determined by \citet{Molster2002b}. That complex shows a sharp rise across the band, furthermore the red-edge of the complex has a strong relatively flat component with few peaks, which is not seen in any of the {\em Spitzer} spectra. We believe that our 28-$\mu $m profiles provide a more accurate/reasonable reflection of the crystalline features present, as these are not affected by the problematic SWS 3D and 3E bands.

\subsubsection{The 33-\mum complex} 

The 33-\mum complex is dominated by the strong forsterite feature at 33.6 $\mu $m, and is particularly prominent in sources with high mass-loss rates. In both the LMC and SMC samples, strong noise around 33 \mum complicates the identification and extraction of features, thus the detection rate may be lower than expected compared to the MW sample. The 33-\mum complex is not seen in the globular cluster stars. 
The complex shows remarkably little variance with metallicity, both in terms of the shape, mean wavelength and FWHM. The mean wavelength of the extracted feature is 33.54 $\mu $m, varying by 0.13 $\mu $m across the three galaxies.  The position of the 33.6-$\mu $m feature is consistent with cool Mg-dominated crystalline olivine grains. If the grains contained iron inclusions the peak of the feature would be shifted to the red \citep{Koike2003, Pitman2010}. Warmer grains would also result in a shift in band position to longer wavelengths \citep{Koike2006}.
The 33-\mum complex also exhibits some less prominent sub-structure. There is a weak feature on the left shoulder of the complex at 32.2 $\mu $m, tentatively identified as diopside (MgCaSi$_{2}$O$_{6}$) by \citet{Molster2002b}. At the red edge of the band we note the presence of an enstatite feature at 35.3 $\mu $m.

\subsubsection{Probable disc sources}

\begin{figure}
  \includegraphics[width=84mm]{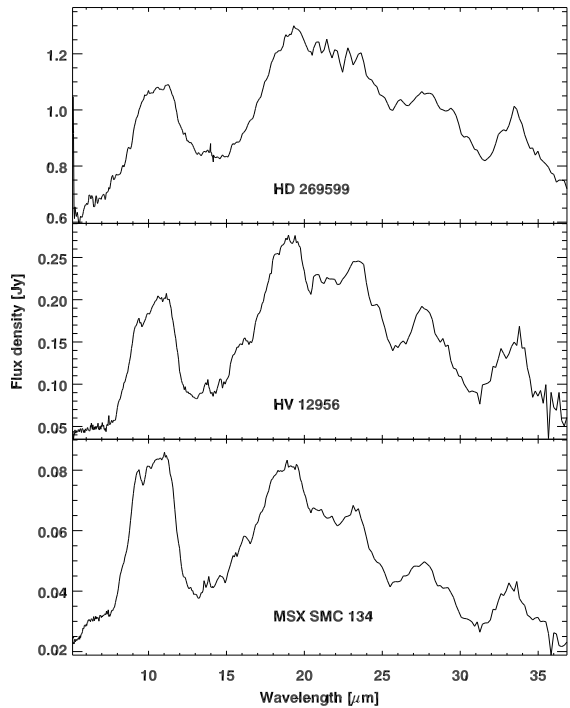}
  \caption{{\em Spitzer} spectra of three sources which display very strong crystalline features. We postulate that these sources have a stable disc-like geometry.}
  \label{Fig:DiscSpec}
\end{figure}

Fig.~\ref{Fig:DiscSpec} shows three sources that exhibit unusually strong crystalline silicate features in their spectra. Strong crystalline silicate features such as these are often found in long-lived circumstellar discs where dust processing creates very high abundances of crystalline silicates \citep{Molster1999, Gielen2008, Gielen2011}. We propose that HD 269599 (SSID 4464), MSX SMC 134 and HV 12956 have a stable disc-like geometry, which invalidates the assumption of a spherically-symmetric outflow for the SED fitting. 
HD 269599 is a member of the young (16 $\pm$ 2 Myr) cluster NGC 1994 \citep{Kumar2008} which contains B and B[e] stars. Because of its large infrared excess it is probable that HD 269599 is also a relatively massive, relatively young star with a disc. Optical spectroscopy confirms that HD 269599 is a B[e] supergiant \citep{Lamers1998, Kastner2010}. 

MSX SMC 134 is RAW 631, an optically identified carbon star. Optical and near-infrared spectroscopy confirm its carbon-rich nature (Sloan et al. in prep, \citealt{vanLoon2008}). The combination of a carbon-rich photosphere and oxygen-rich dust is the hallmark of a silicate carbon star (first discovered by \citealt{LittleMarenin1986}). The leading explanation for these sources is that the silicate dust was trapped in a disc in a binary system before the mass-losing star evolved into a carbon star \citep{LloydEvans1991, Barnbaum1991}.

We could find nothing in the literature that signifies the presence of a disc around HV 12956, but postulate the presence of a disk is required to show these high-contrast crystalline silicate features.

\subsection{The olivine-to-pyroxene ratio}

\begin{figure}
\includegraphics[width=84mm]{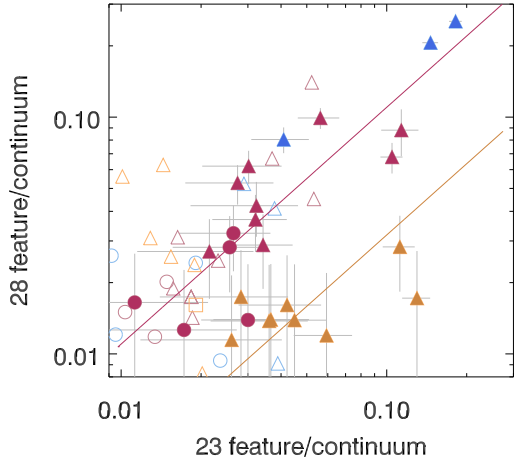}
 \caption{Correlation between the strength of the 23-\mum  feature and the 28-\mum  feature. The symbols are as defined in Fig.~\ref{Fig:MLRvft}, except the filled symbols represent the presence of both a 23- and a 28-\mum feature. The lines show the least square fit to the LMC (red) and MW (orange) features.} 
  \label{Fig:23v28}
\end{figure}

\begin{figure}
\includegraphics[width=84mm]{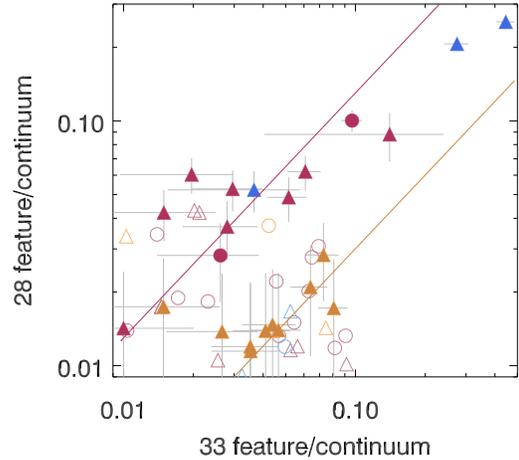}
 \caption{Correlation between the strength of the 33-\mum  feature and the 28-\mum  feature, with best-fit line for each population. The symbols are as defined in Fig.~\ref{Fig:MLRvft}, except the filled symbols represent the presence of both a 33- and a 28-\mum feature. }
  \label{Fig:33v28}
\end{figure}

\begin{figure}
\includegraphics[width=84mm]{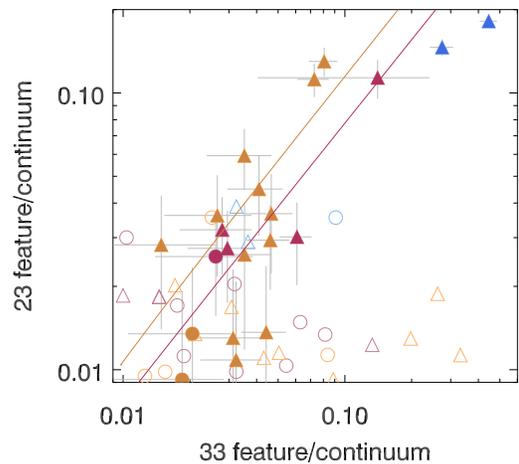}
 \caption{Correlation between the strength of the 23-\mum feature and the 33-\mum feature. The symbols are as defined in Fig.~\ref{Fig:MLRvft}, except the filled symbols represent the presence of both a 23- and a 33-\mum feature.}
  \label{Fig:23v33}
\end{figure}

The variation in spectral shape and the differences between individual bands present in the 28-\mum complexes for the Milky Way, Magellanic Clouds and globular clusters, may indicate a change in the crystalline silicate dust mineralogy with metallicity. 
As the metallicity increases, the enstatite substructure at 28.2 and 29.4 \mum becomes less pronounced, whereas the forsterite substructure at 27.5 \mum becomes more distinct. This dichotomy between the Magellanic Clouds and the Galaxy can be clearly seen in Figs.~\ref{Fig:23v28} and \ref{Fig:33v28} which show the relations between the 23- and 33-\mum forsterite features and the 28-\mum enstatite feature strength. The separation between the populations, as indicated by the difference between the lines of best fit to the crystalline silicate features in each galaxy, suggests a change in crystalline dust composition with metallicity. 
A line of best fit is not plotted for the SMC sources due to insufficient data points. Note that the globular cluster sources do not feature in these plots as they only exhibit the 28-\mum enstatite feature, also indicating a change in dust composition with metallicity. 
This trend is almost certainly a metallicity effect, and not due to changes in mass-loss rate or luminosity as we are sensitive to crystalline silicate features in AGB and RSG sources across several dex in both luminosity and mass-loss rate. 

Figure \ref{Fig:23v33} shows the relation between the 23- and 33-\mum features which have the same primary carrier (forsterite). There is no clear separation between the different metallicity populations. However, several of the Milky Way sources are not very well represented by the best fit line. These sources are experiencing intense mass-loss rates consequently the 23-\mum feature might experiencing self-absorption. 

\subsection{Pulsation period and crystalline feature strength}

\begin{figure*}
  \includegraphics{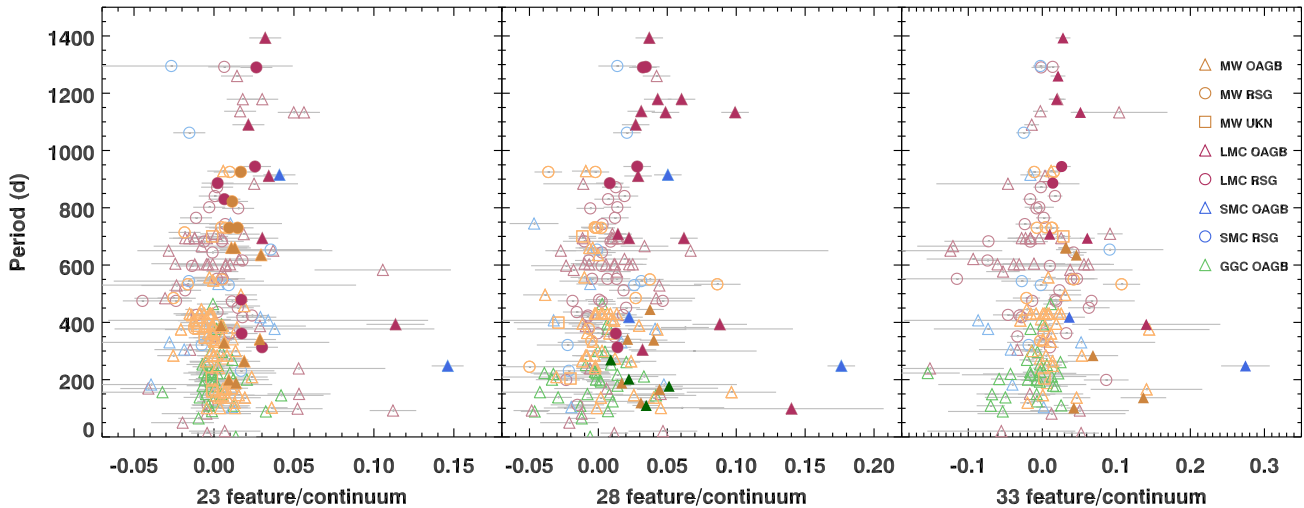}
  \caption{Comparison between pulsation period and the crystalline silicate feature strength. Symbols are the same as Fig.~\ref{Fig:MLRvft}.}
  \label{Fig:Pvft}
\end{figure*}

In general, as an AGB star evolves, the stellar radius increases, it becomes brighter, and its pulsation period increases. A similar period--luminosity relation exists for RSGs, although they often have smaller amplitudes and less regular periodicities. Fig.~\ref{Fig:Pvft} compares the crystalline band strength with pulsation period. Care needs to be taken when comparing trends with pulsations across a metallicity range, since a star's effective temperature is metallicity-dependent and hence so are radius and pulsation period \citep{Wood1990}. We therefore consider each population separately. 
We find that the fraction of sources showing crystalline silicate features is greater for sources with pulsation period longer than 700 days than for sources with short pulsation periods.  Furthermore, in these long period pulsators the strength of the crystalline features appears to be increasing with period. This effect is strongest for the 28 \mum feature in Fig.~\ref{Fig:Pvft}. These stars are typically more evolved with higher dust mass-loss rates than the sources with periods less than 700 days. 

\section{Discussion}\label{sec:discussion}

\subsection{Mass-loss rates}

Determining total mass-loss rates requires an assumption about the scaling relation between the dust and gas in the envelope of AGB stars; this correction factor is not well known, even for Milky Way sources. It was proposed by \cite{vanLoon2000} that the gas-to-dust ratio for O-rich sources is approximately linearly dependant on the initial metallicity. Although this relation has an inherent scatter, it is commonly applied when deriving gas mass-loss rates. Our estimates for the total mass-loss rates are similarly affected by the large uncertainties in the correction factor, which may blur the effects of the gas density on the onset of crystallinity in the envelopes of evolved stars. Furthermore, the precision of the individual gas mass-loss rates will be affected by the assumption that objects in each galaxy are of the same metallicity. Total mass-loss rates, determined from CO emission, may provide a better constraint of the gas density. This would require observations with the full ALMA array for objects at the distance of the Magellanic Clouds (Woods et al. 2012, MNRAS, in press), however, the correlation between CO and total gas mass is also uncertain.

\subsection{Onset of crystallinity} 

Theoretical dust condensation studies predict that the dust \citep{Sogawa1999} or gas \citep{Tielens1998, GailSed1999} column density in the circumstellar envelope is critical for the formation of crystalline silicates. Our observations show that, at high dust densities (${\dot{M}_{dust}}$ $>$ $3 \times 10^{-8}\, {\rm M_{\odot}\, yr^{-1}}$), the majority of the sources in our sample exhibit silicate grains with a crystalline component. However, we first detect evidence for crystallinity at a low dust column density corresponding to ${\dot{M}_{dust}}$ $\sim$ $5 \times 10^{-10}\, {\rm M_{\odot}\, yr^{-1}}$. Approximately 10 per cent of the sources with dust mass-loss rates between $10^{-9.3}$ and $10^{-8}\, {\rm M_{\odot}\, yr^{-1}}$ show the distinctive structure of crystalline silicate bands. A transition in the silicate structure is seen when dust mass-loss rates of $10^{-8}$ to $10^{-7}\, {\rm M_{\odot}\, yr^{-1}}$ are reached. Here, the silicate grains undergo a transition from only a few per cent of sources displaying crystallinity to over 80 per cent of the sources, in less than 1 dex in dust mass-loss rates.  This transition appears to be sharper for the dust mass-loss rates compared to the gas mass-loss rates, although the uncertainties are greater for the gas mass-loss rates. Although tentative, this suggests that crystallinity is predominantly correlated to the dust mass-loss rate, and thus annealing of amorphous silicate grains by radiation is probably the primary formation mechanism for crystalline silicates in the outflows of AGB and RSG stars.

Alternatively, \cite{Sloan2008} argue that in low-density winds grain growth occurs over a long period, allowing the atoms to arrange themselves in the most energetically-favourable lattice structure before being locked in place as the grain accumulates more layers. At higher densities grains will accrete material faster, preventing the migration of atoms. If this was the case one would most likely expect to see a bimodal distribution of crystalline sources, which is not supported by our results.

For sources undergoing the most extreme mass-loss rates a combination of crystallisation processes may be in effect. Here conditions are favourable for both thermal annealing of the amorphous grains and crystalline grains forming by direct condensation from the gas phase, where the higher densities in the dust-formation zone raise the silicate condensation temperature above the glass temperature \citep{Lodders1999, GailSed1999}. This combined effect may in part explain why the sources with high  mass-loss rates tend to have stronger crystalline features.

\subsection{Mineralogy and metallicity}

Our results provide strong evidence for a change in crystalline silicate production with metallicity; here we explore some of the possible reasons for these differences. 

The molecules available for dust production in O-rich stars are limited by the abundances of heavy elements which are not produced on the AGB. The abundance of Mg, Si, Ca, Al, and Fe in the photosphere reflect the initial abundances of the molecular cloud from which the star formed, while the surface abundances of C and O are altered by dredge-up and hot bottom burning. Dredge-up only increases the [O/H] abundance by small amounts. However efficient hot bottom burning in low-metallicity, high-mass AGB stars may result in [O/H] depletion \citep{Ventura2002}. For stars with solar-like abundances, both Mg and Fe are more abundant than Ca and Al (Mg/Fe being approximately unity). These abundances change with overall metallicity. For non-$\alpha$-capture products like Al, this effect is even more severe \citep{Wheeler1989}. 

For solar-metallicity stars, the classical dust condensation sequence in the outflows of O-rich evolved stars predicts that aluminium- or calcium-rich dust grains form first, at temperatures around 1400 K, before silicon and magnesium condense into silicates at slightly lower temperatures \citep{Tielens1990, Lodders1999, GailSed1999}. A parallel condensation sequence, involving magnesium and silicon, starts with the direct condensation of forsterite from the gas phase at temperatures around 1050 K. This is subsequently transformed into enstatite through reactions between forsterite and SiO$_{2}$ gas, according to the equilibrium reaction: 

\begin{equation} \label{eq:ForstEnstEq}
  {\rm 2MgSiO_{3}} \rightleftharpoons {\rm Mg_{2}SiO_{4} + SiO_{2}.}
\end{equation} 

Dust condensation for equilibrium conditions has been suggested to follow the sequence below:

\begin{tabular}{@{}cc@{}}
                                        &                              	    	  \\
${\rm Al_{2}O_{3}, Alumina}$            &                              	          \\	 
$\Downarrow $                           &                              	    	  \\
${\rm Ca_{2}Al_{2}SiO_{7}, Gehlenite}$  &                              	          \\ 	 
$\Downarrow $                           &                              	    	  \\
${\rm CaMgSi_{2}O_{6}, Diopside}$       & 	${\rm Mg_{2}SiO_{4}, Forsterite}$ \\
$\Downarrow $                           &   	$\Downarrow $        	      	  \\
${\rm MgAl_{2}O_{4}, Spinel}$           & 	${\rm MgSiO_{3}, Enstatite}$      \\
$\Downarrow $                           &  	$\Downarrow $       	       	  \\
${\rm CaAl_{2}Si_{2}O_{8}, Anorthite}$ 	&  ${\rm (Mg,Fe)_{2}SiO_{4}, Olivine}$    \\
                                        &                                     	  \\
\end{tabular} 

The condensation sequence in O-AGB stars is slightly different from that of supergiants \citep{Speck2000, Verhoelst2009}, however, the end-point for both sequences results in the formation of magnesium-rich silicates. 
The lower detection rate of crystalline features in RSGs may be a consequence of this alternate condensation pathway, due to the delayed production of Mg-rich amorphous silicates. We speculate that this production is a necessary precursor for the formation of forsterite and enstatite, as it provides suitable material from which to anneal crystalline silicates.

The crystalline silicates in our sample are comprised of Mg-rich olivines and pyroxenes (forsterite: Mg$_{2}$SiO$_{4}$ and enstatite: MgSiO$_{3}$). 
Recently, the crystalline features in the {\em ISO} spectra of  RX Lac, T Cep, T Cet and R Hya were found to be better matched with the iron-rich silicate: Mg$_{0.18}$Fe$_{1.82}$SiO$_{4}$ \citep{Pitman2010,Niyogi2011}. 
These sources occur early on in the silicate emission sequence (which characterises the shape of the 10 \mum complex; see \citealt{Sloan1995}) and lack the classical emission/absorption features which have been attributed to amorphous silicate dust, they also tend to have optically thinner dust shells. However, {\em Herschel} observations of the 69 $\mu$m crystalline silicate band in more evolved AGB stars do not find evidence for iron in the lattice structure of the crystalline olivine \citep{deVries2011}. The sources in our sample lie later on the in the sequence, these tend to have optically thicker shells and a greater abundance of amorphous silicates. Consequently, Fe-rich crystalline silicates may only form early on in the AGB, in low-density winds.

 At solar metallicities, forsterite is the dominant crystalline species, but as the metallicity drops enstatite becomes preferentially formed. We consider five possibilities for this difference below. 

\begin{enumerate}

\item The elemental abundance of Mg, Si and O during dust condensation will also influence the forsterite-to-enstatite ratio. If either Mg or O are depleted then enstatite could preferentially form. One might expect the abundances of Si and Mg to scale from the solar values by the same amount, in which case a lower relative abundance of O would be required for enstatite production. This could be due to water formation locking up free oxygen, changes in the natal C/O ratio with metallicity, or efficient hot bottom burning in low-metallicity high-mass AGB stars.

 In more metal-poor galaxies, dredge-up of newly produced carbon has a greater effect on the C/O ratio in the photosphere of the star \citep{Lattanzio2003}. A higher abundance of CO reduces the the amount of free oxygen available for dust production, which might result in the preferential formation of grains with a pyroxene stoichiometry as they require less oxygen.

The chemical abundances in massive AGB stars can also be modified by hot bottom burning. If the temperature's are high enough (T $> 3 \times 10^6$ K) for the NO cycle and Ne-Na cycle to become active, oxygen will become depleted and sodium should be enhanced \citep{Caciolli2011}.
Oxygen depletion (and associated sodium enhancement) has been observed in several low metallicity  globular cluster stars \citep{Kraft1997, Gratton2001}. These stars have insufficient mass to have undergone hot bottom burning, however, it may be possible that the abundances of the primordial gas was enriched in He, C, N and Na, but depleted in O  by an early population of massive AGB stars \citep{Ventura2001, Decressin2007}.

\item Differences in the temperature gradient in the outflows of O-AGB and RSG stars in the Galaxy, Magellanic Clouds and Galactic globular clusters, will alter the position of equilibrium in the forsterite/enstatite reaction. 
As the temperature of the circumstellar envelope decreases, the position of equilibrium in Eq.~\ref{eq:ForstEnstEq} will shift to favour the exothermic reaction, in this case the production of enstatite. Consequently the ${\rm MgSiO_{3}}$/${\rm Mg_{2}SiO_{4}}$ ratio would increase. This reaction requires that equilibrium is reached.

\item Alternatively, non-equilibrium conditions during dust condensation may alter the order that olivines and pyroxenes form. In denser environments, the dust shells are more likely to reach equilibrium due to the increased probability of interactions with other atoms/molecules, thus for Galactic sources where the partial pressure of dust is greatest, the circumstellar envelopes are more likely to reach equilibrium than sources at lower metallicities. 
 Under equilibrium conditions it is predicted that forsterite will form before enstatite.  If forsterite condenses first then Mg will deplete at twice the rate of Si. Conversely, if enstatite condenses first the [Mg/Si] ratio will remain constant until the Si is exhausted. 

\item Furthermore, the  higher partial pressure of dust-forming elements in the envelopes of Milky Way sources likely results in the growth of larger grains. The size of forsterite grains may also be an important factor in enstatite production: for large grains the surface area per unit volume decreases, reducing available reaction sites for the infusion of ${\rm SiO_{2}}$ in the crystalline lattice.  This curtails the amount of forsterite available for conversion to enstatite, and may result in large forsterite grains coated by a enstatite mantle. 

\item Nucleation of Mg, SiO and H$_2$O in the conditions relative to stellar outflows could also plausibly form enstatite  \citep{Goumans2012}. For silicates to nucleate from gas phase molecules, they must follow a series of thermodynamically- favourable reaction pathways, resulting in a small metastable silicate cluster with an enstatite stoichiometries.  Once a cluster has formed these must grow under appropriate temperature conditions to become macroscopic particles \citep{Gauger1990}. The subsequent growth of this cluster is extremely exothermic, this increase in internal grain temperature may be sufficient for the partial crystallisation of the forming dust particle. Why this nucleation pathway would be favoured in the metal-poor environments of the Magellanic Clouds remains unclear. 

\end{enumerate}

\section{Conclusion}\label{sec:conclusion}

We have analysed \emph{Spitzer} and \emph{ISO} infrared spectra of 217 O-AGB and 98 RSG stars in the Milky Way, Magellanic Clouds and Galactic globular clusters, to explore the onset of crystallinity and investigate how the mineralogy depends on the physical and chemical conditions of the star's envelope. Dust mass-loss rates were established through spectral energy distribution fitting with the {\sc grams} model grid and the mineralogy of the crystalline features determined from the spectra. The main results of this study are summarised as follows:

\begin{itemize}
\item We detect crystalline silicates over 3 dex in dust mass-loss rate down rates of $\sim$10$^{-9}$ M$_\odot$ yr$^{-1}$. 
\item Crystalline silicates are more prevalent in higher mass-loss rates objects, though sources undergoing the highest mass loss do not show the 23-\mum forsterite feature. This is due to the poor contrast of the low temperature of the forsterite grains above the continuum and in some cases (self-)absorption of the short wavelength ($\lambda < 25 \mu$m) crystalline silicate features. 
\item The dust mass-loss rate appears to have a greater influence on the crystalline fraction than the gas mass-loss rate. This may suggest that the annealing of amorphous silicate grains by radiation is probably the primary formation mechanism for crystalline silicates in the outflows of AGB and RSG stars. 
\item  O-AGB stars have a higher proportion of sources with crystalline silicates features than RSGs, however, there is little variation in the structure of the crystalline silicate dust for O-AGB and RSG stars. 
\item We report the presence of a newly detected 22.8 $\mu $m emission feature in the spectra of Milky Way AGB and RSG stars.
\item We detect a change in the crystalline silicate mineralogy with metallicity, with enstatite seen increasingly at low metallicity, while forsterite becomes depleted. This variation in the olivine-to-pyroxene ratio can be explained by a number of possible mechanisms.
\end{itemize}

\section*{Acknowledgements}

We thank the referee, A. K. Speck for the constructive comments which have improved this paper. The authors wish to thank AGGM Tielens for inspiring comments and discussions. OJ acknowledges the support of an STFC studentship, and thanks Academia Sinica Institute of Astronomy and Astrophysics (ASIAA) for their hospitality during the completion of part of this work. FK acknowledges support from the National Science Council in the form of grant NSC100-2112-M-001-023-MY3. B.A.S. acknowledges support from NASA ADP NNX11AB06G. This publication makes use of data products from the Two Micron All Sky Survey (2MASS), which is a joint project of the University of Massachusetts and the Infrared Processing and Analysis Center/California Institute of Technology, funded by NASA and the National Science Foundation; data products from the Optical Gravitational Lensing Experiment OGLE-III on-line catalogue of variable stars; the SIMBAD database and the VizieR catalogue access tool, CDS, Strasbourg, France; NASA's Astrophysics Data System (ADS) Bibliographic Services.


\input{journaldefs}

\bibliographystyle{aa}
\bibliography{libby}


 \end{document}

%% file: journaldefs.tex

\def\aj{AJ}					
\def\actaa{Acta Astron.}                        
\def\araa{ARA\&A}				
\def\apj{ApJ}					
\def\apjl{ApJL}					
\def\apjs{ApJS}					
\def\ao{Appl.~Opt.}				
\def\apss{Ap\&SS}				
\def\aap{A\&A}					
\def\aapr{A\&A~Rev.}				
\def\aaps{A\&AS}				
\def\azh{AZh}					
\def\baas{BAAS}					
\def\jrasc{JRASC}				
\def\memras{MmRAS}				
\def\mnras{MNRAS}				
\def\pra{Phys.~Rev.~A}				
\def\prb{Phys.~Rev.~B}				
\def\prc{Phys.~Rev.~C}				
\def\prd{Phys.~Rev.~D}				
\def\pre{Phys.~Rev.~E}				
\def\prl{Phys.~Rev.~Lett.}			
\def\pasp{PASP}					
\def\pasj{PASJ}					
\def\qjras{QJRAS}				
\def\skytel{S\&T}				
\def\solphys{Sol.~Phys.}			
\def\sovast{Soviet~Ast.}			
\def\ssr{Space~Sci.~Rev.}			
\def\zap{ZAp}					
\def\nat{Nature}				
\def\iaucirc{IAU~Circ.}				
\def\aplett{Astrophys.~Lett.}			
\def\apspr{Astrophys.~Space~Phys.~Res.}		
\def\bain{Bull.~Astron.~Inst.~Netherlands}	
\def\fcp{Fund.~Cosmic~Phys.}			
\def\gca{Geochim.~Cosmochim.~Acta}		
\def\grl{Geophys.~Res.~Lett.}			
\def\jcp{J.~Chem.~Phys.}			
\def\jgr{J.~Geophys.~Res.}			
\def\jqsrt{J.~Quant.~Spec.~Radiat.~Transf.}	
\def\memsai{Mem.~Soc.~Astron.~Italiana}		
\def\nphysa{Nucl.~Phys.~A}			
\def\physrep{Phys.~Rep.}			
\def\physscr{Phys.~Scr}				
\def\planss{Planet.~Space~Sci.}			
\def\procspie{Proc.~SPIE}			
\let\astap=\aap
\let\apjlett=\apjl
\let\apjsupp=\apjs
\let\applopt=\ao
